\documentclass[lettersize,journal]{IEEEtran}
\usepackage{amsmath,amsfonts}
\usepackage{algorithmic}
\usepackage{algorithm}
\usepackage{array}
\usepackage[caption=false,font=normalsize,labelfont=sf,textfont=sf]{subfig}
\usepackage{booktabs}
\usepackage{float}
\usepackage{authblk}
\usepackage{CJKutf8}
\usepackage{amssymb}
\usepackage{stfloats}
\usepackage{url}
\usepackage{verbatim}
\usepackage{graphicx}
\usepackage{cite}
\usepackage{comment}
\DeclareUnicodeCharacter{2212}{\ensuremath{-}}

\usepackage[most]{tcolorbox}
\usepackage{xcolor}
\usepackage{tikz}

\usepackage[normalem]{ulem}

\definecolor{posStrong}{HTML}{C0392B}  
\definecolor{posMild}{HTML}{E67E22}    
\definecolor{neutral}{HTML}{7F8C8D}    
\definecolor{negMild}{HTML}{2980B9}    
\definecolor{negStrong}{HTML}{1A5276}  
\definecolor{wrapperbg}{HTML}{F4F6F7}

\begin{document}

\title{CHARM: Charge Calibration and Acoustic Rescue \\ for LLM-based Multimodal Sarcasm Detection}

\author{
Qiyang Sun \IEEEmembership{Student Member, IEEE},
Yi Chang*, Yupei Li, 
Xi Shao \IEEEmembership{Member, IEEE},
Zixing Zhang*
\IEEEmembership{Senior Member, IEEE}, 
and Björn W.\ Schuller \IEEEmembership{Fellow, IEEE}

\thanks{Qiyang Sun, Yi Chang, and Yupei Li are with GLAM -- the Group on Language, Audio, \& Music, Imperial College London, UK. e-mail: q.sun23@imperial.ac.uk; yichang312@gmail.com; yupei.li22@imperial.ac.uk}

\thanks{Xi Shao is with the College of Telecommunications and Information Engineering, Nanjing University of
Posts and Telecommunications, China, e-mail: shaoxi@njupt.edu.cn}

\thanks{Zixing Zhang is with the College of
Computer Science and Electronic Engineering, Hunan University, China; Zixing Zhang is also with the Shenzhen Research Institute, Hunan University, China. e-mail: zixingzhang@hnu.edu.cn}

\thanks{Björn W.\ Schuller is with GLAM -- the Group on Language, Audio, \& Music, Imperial College London, UK; CHI -- Chair of Health Informatics, TUM University Hospital, Germany; relAI -- the Konrad Zuse School of Excellence in Reliable AI, Germany; MDSI -- Munich Data Science Institute, Germany; and MCML -- Munich Center for Machine Learning, Germany. e-mail: bjoern.schuller@imperial.ac.uk}
\thanks{Corresponding author: Yi Chang, Zixing Zhang}
\thanks{Manuscript received April 19, 2021; revised August 16, 2021.}}

\markboth{Journal of \LaTeX\ Class Files,~Vol.~14, No.~8, August~2021}%
{Shell \MakeLowercase{\textit{et al.}}: A Sample Article Using IEEEtran.cls for IEEE Journals}


\maketitle

\begin{abstract}

Sarcasm detection, the identification of discrepancies between literal and intended meaning, is a fundamental task in affective computing. However, zero-shot instruction-tuned Large Language Models (LLMs) systematically over-predict the positive (sarcastic) class across the entire capability spectrum, while the prosodic cues humans rely on remain underexploited and transfer unevenly across languages. We introduce \textbf{CHARM} (\textbf{Ch}arge Calibration and \textbf{A}coustic \textbf{R}escue for \textbf{M}ultimodal Sarcasm Detection), a training-free framework that couples two modules. \textbf{Bi}directional \textbf{C}harge \textbf{C}alibration (\textbf{BiCAL}) steers the LLM toward opposing sarcastic and literal verdicts along a symmetric axis of charged prompts; the induced directional biases cancel by construction, and a simple aggregation recovers an unbiased pragmatic signal. \textbf{A}coustic \textbf{L}ate-\textbf{F}usion \textbf{R}escue (\textbf{ALFR}) then fuses the calibrated votes with prosodic descriptors and LLM-generated auditory-perception probes through a shallow classifier, actively down-weighting saturated text votes in favour of acoustic evidence. Without fine-tuning any backbone, BiCAL attains the highest reported zero-shot text-only Macro-F1 of 0.787 on MUStARD, while ALFR lifts weak backbones by up to +0.382 Macro-F1 on CMMA. A Stouffer meta-analysis confirms statistical significance on MUStARD and CMMA ($Z = 13.89$ and $Z = 34.64$, respectively; $p < 10^{-43}$). Our analysis further uncovers a cross-cultural prosodic decoupling: low-level acoustics fail to transfer across languages, whereas high-level perceptual abstractions remain robust. Together, these components yield an explainable, cross-lingual multimodal detector.

\end{abstract}


\section{Introduction}
\IEEEPARstart{H}{umans} recognise sarcasm almost instantly from a dry tone or a knowing pause. However, contemporary instruction-tuned Large Language Models (LLMs) systematically err in the opposite direction. Sarcasm detection involves identifying the deliberate mismatch between the literal and intended meaning of a speaker~\cite{santosa2025sarcasm}. This operation serves as a cornerstone task within affective computing. This field aims to enable machines to perceive, understand, interpret, and express human emotions~\cite{picard1997affective, sun2026towards}. Unlike standard sentiment analysis~\cite{amiriparian2024muse} or emotion recognition~\cite{li2025gatedxlstm}, sarcastic utterances rarely carry explicit sentiment cues; instead, they hinge on implicit lexical and prosodic nuances. Consequently, automated systems must jointly analyse lexical content, dialogue context, and prosody to infer intent~\cite{gao2025spoken}. Whereas humans process these multimodal signals effortlessly, automated systems still struggle to interpret them reliably~\cite{dhumpati2025enhancing}, opening a costly performance gap. Accurate sarcasm detection nonetheless remains crucial for conversational AI, intelligent customer service, and social-media analysis~\cite{mai2024llama}, where misinterpreting sarcasm yields inappropriate responses that erode both operational effectiveness and user trust.

Recently, large language models (LLMs) have become increasingly prominent in affective computing, and various efforts leverage their emergent reasoning capabilities for subjective pragmatic tasks~\cite{zhang2024refashioning, zhang2026affective}. When executing such tasks, however, instruction-tuned LLMs exhibit a pronounced sycophancy bias~\cite{perez2023discovering}: reinforcement learning from human feedback (RLHF) routinely rewards alignment with explicit user presuppositions rather than objective neutrality~\cite{sharma2024towards}. A presupposition is the stance implicitly encoded in the framing of a query; merely asking ``Is the speaker being sarcastic?'' already foregrounds sarcasm as a live possibility, and a sycophantic model treats this framing as a direct cue to affirm rather than evaluating the input as a neutral proposition to adjudicate. The key observation is that this bias is \emph{directional}: if it behaves like a signed ``charge'' that pushes the verdict toward one pole, then symmetrically applying opposing positive and negative prompt charges should let the biases cancel and expose the underlying signal -- the intuition we later formalise as charge calibration.

This directional skew emerges clearly within recent benchmarks. Extant studies systematically test off-the-shelf models on sarcasm detection~\cite{mai2024llama,11146812}. These evaluations demonstrate that LLMs remain highly sensitive to the precise prompting formulation~\cite{11146812}. Few-shot exemplars provide only moderate performance gains. Furthermore, this strategy heavily penalises computational efficiency by exponentially increasing inference latency~\cite{mai2024llama}. Heavier prompt engineering remedies encompass both few-shot exemplars and elaborate chain-of-thought scaffolds. Although these extensive interventions reduce prompt sensitivity, they fail to eliminate it entirely~\cite{11146812}. Parallel multimodal frameworks add visual instructions and dense-retrieval networks~\cite{tang2024leveraging}, yet concentrate almost exclusively on textual or visual modalities. Consequently, the acoustic and prosodic cues of spoken sarcasm remain comparatively underexplored. The problem sharpens
under strict zero-shot constraints, where external retrieved exemplars are
unavailable: zero-shot LLM classifiers are already prone to systematic label
bias~\cite{zhao2021calibrate}, and here alignment-induced answer
biases~\cite{sharma2024towards} drive severe
positive-class over-prediction across backbones. Complex reasoning prompts do not uniformly mitigate this issue but instead introduce directional biases of their own ~\cite{turpin2023language}, and neither generic CoT prompting ~\cite{11146812} nor structured step-by-step variants ~\cite{yao2025sarcasm} robustly eliminate the residual skew. Zero-shot text-only performance thus remains bottlenecked by systematic prompt skew, motivating a method that targets the underlying bias directly instead of constructing ever-larger prompts.

The integration of the acoustic modality represents a natural remediation for textual bias. However, this multimodal expansion introduces a secondary, orthogonal gap. Large Audio-Language Models (LALMs) offer a viable path for affective computing~\cite{chu2023qwen} and perform well on standard speech emotion recognition~\cite{yang2025towards}, but their training distributions lean heavily on literal semantic alignment such as automatic speech recognition~\cite{radford2023robust}, leaving fine-grained pragmatic anomalies like sarcasm underexplored within these foundation models. Native end-to-end Omni models attempt to unify speech and text processing~\cite{xu2025qwen2} yet suffer a severe text-dominance bias in their deep layers~\cite{wu2025language}: under holistic querying, the textual stream overrides the acoustic evidence, and the black-box layers fail to decouple the conflicting intents carried by text and prosody~\cite{agarwal2025rethinking}. Such end-to-end fusion therefore ignores fine-grained raw acoustic cues during binary inference~\cite{chen2026audio} and renders the decision an opaque, unexplainable black box~\cite{sun2025explainable,10847866}. This acoustic blindness is orthogonal to the prompt-induced bias above, since averaging symmetric prompts cannot recover a signal the model never attended to. Although fine-tuning presents an alternative, it remains impractical in this context for three reasons. First, annotated audio-text sarcasm data are scarce and heavily class-imbalanced; thus, adapting billion-parameter architectures risks overfitting. Second, the strongest commercial LLMs are restricted behind inference APIs and cannot be fine-tuned. Finally, models optimised for one language's sarcastic markers transfer poorly across cultures. These constraints call for a structural remedy that keeps the foundation models frozen. It elicits prosodic evidence explicitly and reliably, rather than trusting a single end-to-end verdict.

To address these two bottlenecks, we introduce a unified framework, Charge Calibration and Acoustic Rescue for Multimodal Sarcasm Detection (\textbf{CHARM}), that decouples pragmatic reasoning into two independent yet complementary components. The first is Bidirectional Charge Calibration (\textbf{BiCAL}), which targets the over-positivity bias at the purely textual level. Rather than relying on a single prompt, BiCAL deploys symmetric prompt variants along the charge axis introduced above, so the native alignment bias cancels during zero-shot inference; an ensemble voting scheme then aggregates these configurations into a stable text-based prediction.

The second component is the Acoustic Late-Fusion Rescue (\textbf{ALFR}) model, which re-targets the multimodal Omni architecture toward its inherent processing strengths. ALFR employs one Omni LLM to extract four multi-class paralinguistic probes and openSMILE~\cite{eyben2010opensmile} to extract fine-grained raw physical acoustic features; a lightweight classification head then integrates these signals with the calibrated text predictions from BiCAL. This late-fusion routing affords high feature-level explainability and rescues weaker language models from single-class text-based collapse. We evaluate the framework extensively on the English MUStARD~\cite{castro2019towards} and Chinese CMMA~\cite{zhang2023cmma} datasets.

The main contributions of this work are summarised as follows:

\begin{itemize}
\item \textbf{Bidirectional Charge Calibration (BiCAL):} A zero-shot text-only calibration method that neutralises RLHF-induced positivity bias through symmetric prompt-charge ensembling. BiCAL yields a consistent Macro-F1 lift across all seven LLM backbones on both MUStARD and CMMA benchmarks, delivering Stouffer significance scores of $Z = 5.85$ and $Z = 19.29$, respectively. It attains the highest reported zero-shot, text-only Macro-F1 of 0.787 on MUStARD with DeepSeek-V4-Pro, outperforming prior training-free prompting methods, and establishes the first zero-shot LLM benchmark on the CMMA corpus.

  \item \textbf{Acoustic Late-Fusion Rescue (ALFR):} A lightly supervised multimodal extension that fuses BiCAL votes, low-level openSMILE prosodic statistics, and frozen audio-language-model perception probes through a shallow classifier head. ALFR delivers up to $+0.382$ absolute Macro-F1 rescue on the weakest CMMA backbone, with across-backbone Stouffer meta-analyses of $Z = 13.89$ on MUStARD and $34.64$ on CMMA ($p < 10^{-43}$).

  \item \textbf{Cross-cultural Prosodic Decoupling:} We quantify, for the first time on these benchmarks, a $3.3\times$ collapse of low-level prosodic effect sizes between English MUStARD and Chinese CMMA. Conversely, high-level perception probes strengthen cross-culturally. This performance pattern is consistent with prior cross-linguistic acoustic studies of sarcastic speech. 
\end{itemize}
\section{Related Work}
\subsection{LLMs in Sarcasm Detection}
LLMs demonstrate strong reasoning capabilities, and their application to sarcasm detection has become an active research area. In the text modality, Zhang et al.~\cite{11146812} propose the SarcasmBench benchmark, systematically evaluating various LLMs on pure text sarcasm understanding and highlighting their limitations of current models in complex pragmatic reasoning. In the multimodal domain, Li et al.~\cite{li2025leveraging} utilise LLMs to assist the annotation of speech sarcasm data and construct a large-scale multimodal dataset, PodSarc. Mai et al.~\cite{mai2024llama} apply prompt engineering with varying numbers of few-shot examples to probe Llama 3 on textual sarcasm detection, finding that additional exemplars yield only moderate accuracy gains while non-linearly escalating inference latency and penalising computational efficiency. Tang et al.~\cite{tang2024leveraging} employ visual instructions and demonstration-retrieval augmentation to recast multimodal sarcasm detection as an end-to-end generative task. Furthermore, they adopt parameter-efficient fine-tuning such as Low-Rank Adaptation (LoRA) to train LLMs directly on visual-text sarcasm data.

Despite these advances, two limitations persist. First, few-shot prompting only partially alleviates pragmatic-reasoning bottlenecks: 
it requires numerous contextual exemplars, causing the input token length to grow linearly and the self-attention complexity to increase quadratically, and it depends on high-quality external annotations. Crucially, it fails to resolve the inherent sycophancy bias of LLMs under zero-shot settings, where uncalibrated alignment triggers systematic positive-class over-prediction. Second, existing multimodal frameworks rely on parameter-efficient fine-tuning such as LoRA for modal alignment, which incurs substantial training time and data costs and is inapplicable to closed-source frontier models served behind inference APIs. The literature therefore lacks a lightweight, plug-and-play architecture that keeps the foundation backbones entirely frozen.

\subsection{LALMs and Acoustic Integration}

Recent affective computing research actively extends speech processing through LALMs~\cite{chu2023qwen}. Frameworks such as Qwen-Audio align continuous audio waveforms with textual tokens. This structural alignment establishes a general semantic understanding capability. Consequently, these architectures deliver highly competitive results on standard speech emotion recognition. Contemporary speech LLMs, however, tend to prioritise semantic content and overlook the paralinguistic cues that carry affect, yielding shallow emotional interpretations in complex interactions~\cite{wang2025empathy}. In parallel, historical literature shows that low-level physical acoustic descriptors remain foundational indicators of emotion~\cite{gao2025spoken}; recent studies therefore feed these vocal nuances directly into the prompt, which markedly strengthens LLM-based emotion recognition~\cite{wu2025beyond}.

Despite this progress, LALM research on joint audio-text sarcasm detection remains scarce. Most multimodal baselines target general sentiment categorisation and fail to capture the implicit pragmatic mismatch that defines sarcasm~\cite{gao2025spoken}. End-to-end audio models also face distinctive challenges on complex dialogues: Banerjee et al.~\cite{banerjee2025llms}, for instance, evaluate frontier generative architectures on multi-party interactions and observe severe hallucinations during conversational tracking. Because these models entangle cross-modal fusion within deeply coupled hidden layers, their monolithic black-box design obscures the individual contributions of the text stream and the speech channel. The systematic exploitation of acoustic descriptors thus remains underexplored in the LLM era, particularly the deployment of prosody as an explicit, separable signal for sarcasm detection.

\section{Methodology}
\begin{figure*}[t]
\centering
\includegraphics[width=\textwidth]{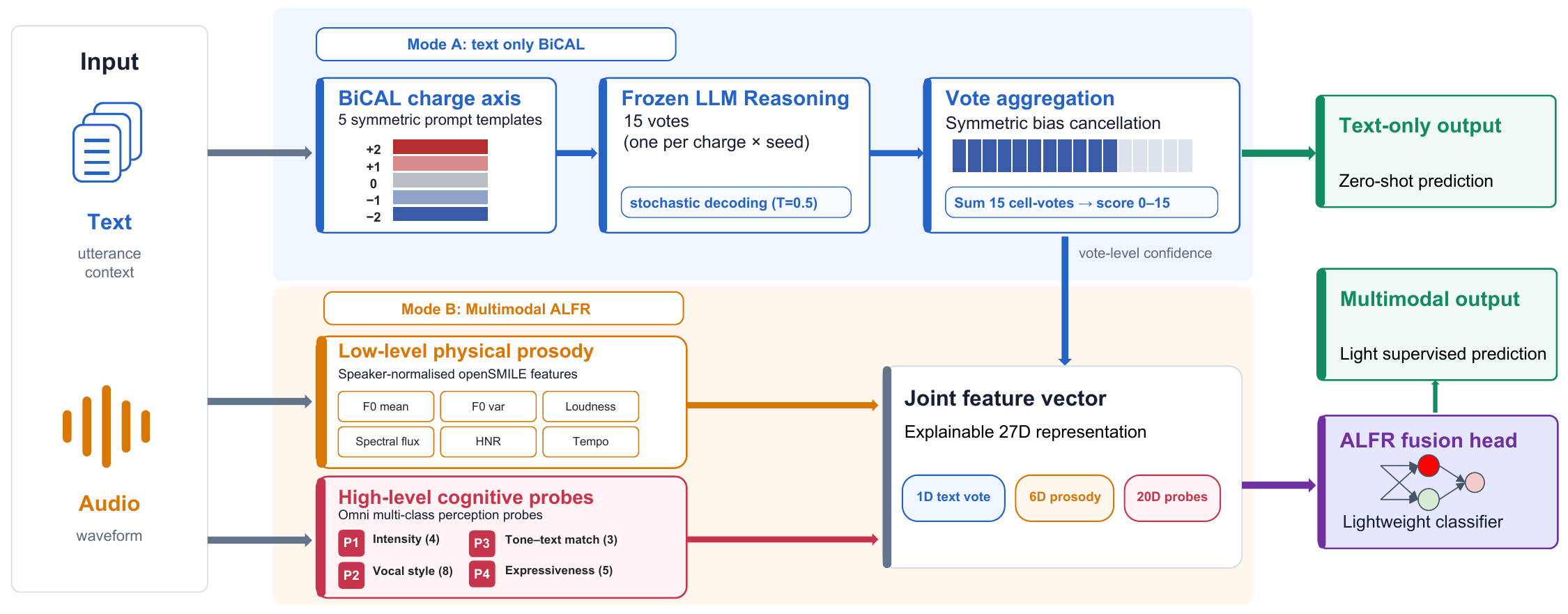}
\caption{Architecture of the \textsc{CHARM} framework. \textbf{Mode A (text-only \textsc{BiCAL}):} Five symmetric charge templates span from $-2$ to $+2$. These configurations elicit binary sarcasm judgements from a frozen LLM via stochastic decoding ($T{=}0.5$). Aggregating predictions across three random seeds yields fifteen cell-level votes. This symmetric layout cancels prior-induced token bias. The pipeline outputs a calibrated pragmatic confidence score within $[0,15]$. \textbf{Mode B (acoustic late-fusion rescue, \textsc{ALFR}):} The framework concatenates the calibrated text vote with multimodal acoustic descriptors. The combined input incorporates six low-level openSMILE $z$-features and twenty high-level Omni perception probes to construct 27 total dimensions. An optimal shallow classifier head resolves the cross-lingual acoustic gap. This integration operates cleanly without parametric weight updates or back-propagation into either foundation backbone.} 
\label{fig:charm_overview}
\end{figure*}

The \textsc{CHARM} framework is built on two functionally orthogonal modules that share neither parameters nor a fused multimodal representation space (cf.\ Figure~\ref{fig:charm_overview}). The \textsc{BiCAL} module operates strictly within the textual domain, neutralising the directional verdict bias of a frozen LLM through a symmetric ensemble of charge-typed prompts. The \textsc{ALFR} module is invoked only when textual evidence saturates. This late-fusion layer fuses the resulting calibrated vote with a compact acoustic feature vector through a shallow per-LLM classifier.

\subsection{Problem Formulation}

We formalise sarcasm detection as a binary classification problem. Let $\mathcal{U}$ denote the space of textual utterances. Let $\mathcal{C} = \mathcal{U}^{k}$ represent the space of length-bounded dialogue contexts. Let $\mathcal{A}$ denote the space of speech-audio waveforms. A multimodal sample is defined as a tuple
\begin{equation}
x = (u, c, a) \in \mathcal{X} = \mathcal{U} \times \mathcal{C} \times (\mathcal{A} \cup \{\varnothing\}),
\end{equation}
where the audio component $a$ may be absent in text-only settings. This configuration is denoted by the null symbol $\varnothing$. The target $y \in \mathcal{Y} = \{0,1\}$ represents the ground-truth pragmatic label. A value of $y=1$ indicates a sarcastic instance. The objective is to construct a decision function that predicts $y$ from $x$.

\subsection{Framework Overview}
The \textsc{CHARM} framework consists of two complementary modules. These components share neither parameters nor a fused representation space (cf.\ Figure~\ref{fig:charm_overview}). The system operates in one of two modes depending on the available modality profile:
\begin{equation}
\label{case}
\
f_{\theta}(x) =
\begin{cases}
\mathbb{1}\bigl[\, v > 7 \,\bigr] & \text{Stage 1 (BiCAL, text-only)}  \\
\mathcal{M}\bigl(x_{\text{joint}}\bigr) & \text{Stage 2 (BiCAL\,+\,ALFR, multimodal). }
\end{cases}
\end{equation}
In Stage 1, the text-only BiCAL module (\S\ref{sec:bical_design}) maps the linguistic input to a calibrated vote count. This operation yields $v \in \{0,\dots,15\}$. The framework thresholds this value directly to generate a prediction.

In Stage 2, the ALFR module (\S\ref{sec:alfr_design}) assembles a 27-dimensional heterogeneous feature vector $x_{\text{joint}}$. The pipeline routes this vector through a shallow classifier $\mathcal{M}$. All foundation backbones, prosody extractors, and perceptual probes remain frozen throughout execution. Only the classifier layer $\mathcal{M}$ consumes task labels. This process occurs exclusively through cross-validation folds. Consequently, \textsc{CHARM} operates as a zero-shot end-to-end system in Stage 1 and a lightly supervised framework in Stage 2. This design positions the method cleanly between purely zero-shot prompting and fully fine-tuned systems.

\subsection{Bidirectional Charge Calibration (BiCAL)}
\label{sec:bical_design}
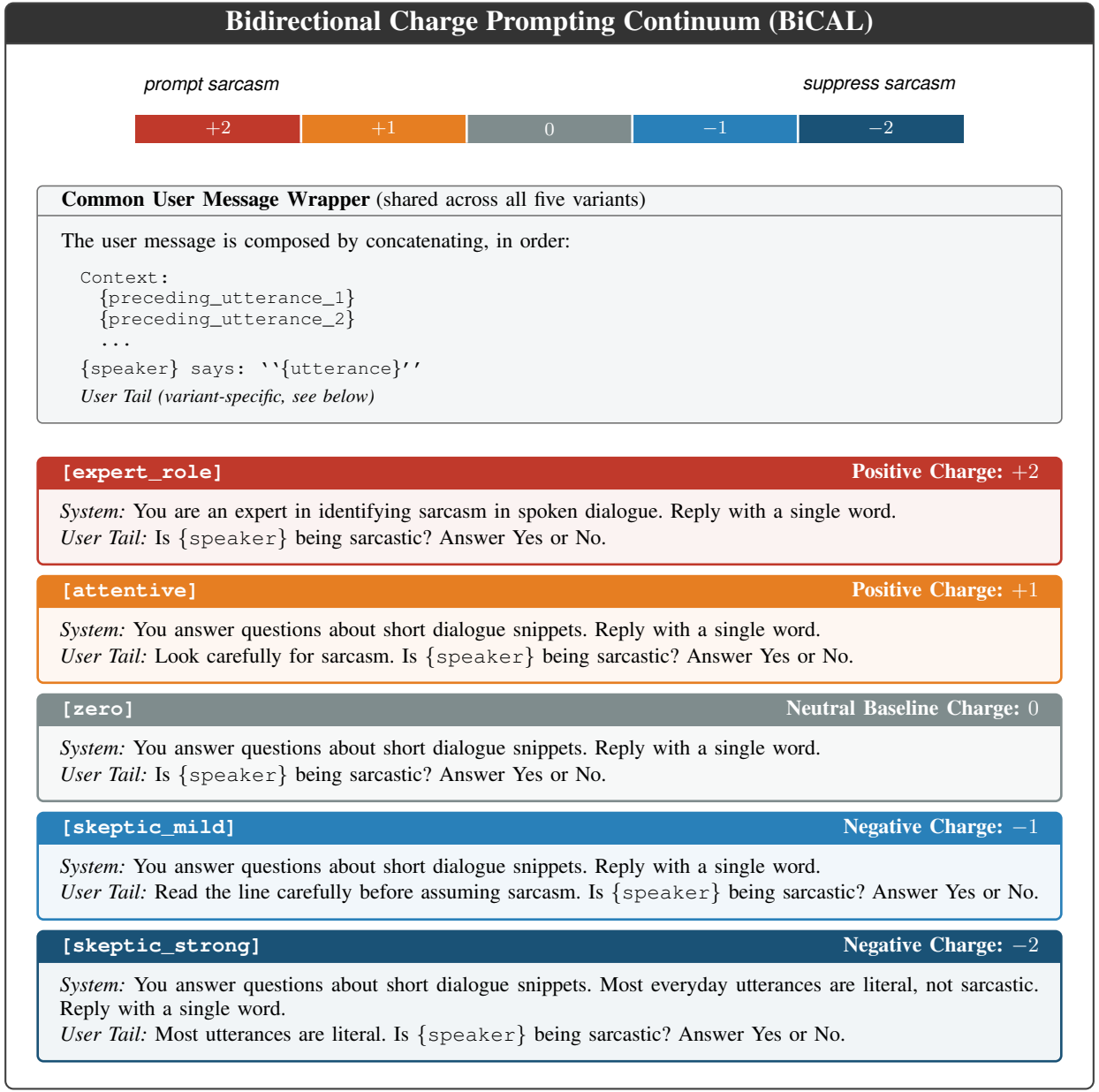
\begin{figure*}[htbp]
\centering

\begin{tcolorbox}[
    enhanced, width=0.9\textwidth,
    colback=white, colframe=black!75, boxrule=0.8pt,
    arc=3pt, left=10pt, right=10pt, top=8pt, bottom=8pt,
    title=\centering\textbf{\large Bidirectional Charge Prompting Continuum (BiCAL)},
    coltitle=white, colbacktitle=black!80, halign title=center,
    fonttitle=\bfseries
]

\begin{center}
\begin{tikzpicture}[font=\footnotesize\sffamily]
  \def\w{12.4}
  \def\seg{2.48} 
  \foreach \i/\col in {0/posStrong, 1/posMild, 2/neutral, 3/negMild, 4/negStrong} {
      \fill[\col] (\i*\seg,0) rectangle ({(\i+1)*\seg},0.42);
  }
  \foreach \i in {1,2,3,4} {
      \draw[white, line width=1pt] (\i*\seg,0) -- (\i*\seg,0.42);
  }
  \foreach \i/\lab in {0/{$+2$}, 1/{$+1$}, 2/{$0$}, 3/{$-1$}, 4/{$-2$}} {
      \node[white,font=\bfseries\footnotesize] at ({(\i+0.5)*\seg},0.21) {\lab};
  }
  \node[anchor=west] at (0,0.85) {\textit{prompt sarcasm}};
  \node[anchor=east] at (\w,0.85) {\textit{suppress sarcasm}};
\end{tikzpicture}
\end{center}
\vspace{2pt}

\begin{tcolorbox}[
    enhanced, colback=wrapperbg, colframe=black!55,
    boxrule=0.6pt, arc=2pt, left=8pt, right=8pt, top=5pt, bottom=5pt,
    title=\textbf{Common User Message Wrapper}\;\textnormal{(shared across all five variants)},
    coltitle=black, colbacktitle=wrapperbg, fonttitle=\small\bfseries,
    boxsep=2pt
]
\small
The user message is composed by concatenating, in order:
\begin{flushleft}\ttfamily\footnotesize
\hspace*{1em}Context:\\
\hspace*{2em}\{preceding\_utterance\_1\}\\
\hspace*{2em}\{preceding\_utterance\_2\}\\
\hspace*{2em}\ldots\\[3pt]
\hspace*{1em}\{speaker\} says: ``\{utterance\}''\\[3pt]
\hspace*{1em}\rmfamily\itshape User Tail (variant-specific, see below)
\end{flushleft}
\end{tcolorbox}

\vspace{4pt}

\newtcolorbox{chargebox}[2]{
    enhanced, colback=#1!6, colframe=#1, boxrule=1pt,
    arc=2pt, left=7pt, right=7pt, top=4pt, bottom=4pt,
    title=#2, coltitle=white, colbacktitle=#1,
    fonttitle=\small\bfseries, boxsep=2pt, before skip=4pt, after skip=4pt
}

\begin{chargebox}{posStrong}{\texttt{[expert\_role]}\hfill Positive Charge: $+2$}
\small\textit{System:} You are an expert in identifying sarcasm in spoken dialogue. Reply with a single word.\\[1pt]
\textit{User Tail:} Is \texttt{\{speaker\}} being sarcastic? Answer Yes or No.
\end{chargebox}

\begin{chargebox}{posMild}{\texttt{[attentive]}\hfill Positive Charge: $+1$}
\small\textit{System:} You answer questions about short dialogue snippets. Reply with a single word.\\[1pt]
\textit{User Tail:} Look carefully for sarcasm. Is \texttt{\{speaker\}} being sarcastic? Answer Yes or No.
\end{chargebox}

\begin{chargebox}{neutral}{\texttt{[zero]}\hfill Neutral Baseline Charge: $0$}
\small\textit{System:} You answer questions about short dialogue snippets. Reply with a single word.\\[1pt]
\textit{User Tail:} Is \texttt{\{speaker\}} being sarcastic? Answer Yes or No.
\end{chargebox}

\begin{chargebox}{negMild}{\texttt{[skeptic\_mild]}\hfill Negative Charge: $-1$}
\small\textit{System:} You answer questions about short dialogue snippets. Reply with a single word.\\[1pt]
\textit{User Tail:} Read the line carefully before assuming sarcasm. Is \texttt{\{speaker\}} being sarcastic? Answer Yes or No.
\end{chargebox}

\begin{chargebox}{negStrong}{\texttt{[skeptic\_strong]}\hfill Negative Charge: $-2$}
\small\textit{System:} You answer questions about short dialogue snippets. Most everyday utterances are literal, not sarcastic. Reply with a single word.\\[1pt]
\textit{User Tail:} Most utterances are literal. Is \texttt{\{speaker\}} being sarcastic? Answer Yes or No.
\end{chargebox}

\end{tcolorbox}

\caption{The Symmetrical Bidirectional Charge Prompting Continuum within the BiCAL Module. A shared \emph{common user message wrapper} feeds all five variants; only the \emph{System} message and \emph{User Tail} vary along the symmetric charge axis, from strongly prompting sarcasm ($+2$) to strongly suppressing it ($-2$). Colour
encodes charge polarity (warm $=$ positive, cool $=$ negative). Dynamic input slots are shown in a typewriter typeface.}
\label{fig:prompt_charges}
\end{figure*}

We introduce Bidirectional Charge Calibration (BiCAL) to neutralise prompt-induced behavioural skews by mapping textual inputs to a calibrated discrete voting profile, without resorting to sequential reasoning chains. Let $\mathcal{C}$ denote a set of five textual charge templates that balance stance-forcing instructions across symmetric pragmatic poles; Figure~\ref{fig:prompt_charges} gives their exact typographic layout and lexical boundaries.

\subsubsection*{Mathematical Rationale of Bias Cancellation}
We provide a theoretical justification for this layout. Let $L(u, c, p)$ denote the latent predictive log odds emitted by the backbone under a template $p \in \mathcal{P}$ for a text tuple. We formalise this conditional log odds function as
\begin{equation}
L(u, c, p) = \ln \frac{P(\hat{y}=1 \mid u, c, p)}{1 - P(\hat{y}=1 \mid u, c, p)}.
\end{equation}
Due to human preference alignment constraints, the predictive distribution suffers from directional bias. We decompose this latent log odds function into two main components
\begin{equation}
L(u, c, p) = f(u, c) + \delta(p),
\end{equation}
and the variable $f(u, c)$ represents the unbiased pragmatic target signal. The variable $\delta(p)$ represents the additive prompt specific bias. Positive charge templates generate an upward inflation where $\delta(p_+) > 0$. Negative charge templates impose a downward compression where $\delta(p_-) < 0$. 

Our framework maps these continuous variables to discrete binary outputs via a monotonic logistic mapping followed by hard thresholding. Since the logistic sigmoid is strictly monotonic and centrosymmetric around the origin, our symmetric prompt layout enforces a zero-sum equilibrium over the bias space
\begin{equation}
\sum_{p \in \mathcal{C}} \delta(p) = 0 .
\end{equation}
Because the bias perturbations are distributed symmetrically around the true signal, the downstream majority voting protocol acts as a non-parametric median estimator. This aggregation cancels out the opposing directional skews in the final probability domain. This mathematical structure isolates the unbiased target representation without requiring direct logit access during decoding.

\subsubsection*{Nested Voting Framework and Decision Paths}
Let $\mathcal{S}$ represent a set of three independent random seeds, so that the ensemble spans fifteen template--seed configuration pairs. For each pair of template $i$ and seed $j$, the language model generates three response samples under stochastic decoding, implementing a localised self-consistency protocol. Let $o_{i,j,k} \in \{0, 1\}$ denote the parsed binary output of the $k$-th response sample. We resolve the local consolidated vote $\phi_{i,j}$ through majority thresholding
\begin{equation}
\phi_{i,j} = \mathbb{1}\left[ \, \sum_{k=1}^{3} o_{i,j,k} \;\ge\; 2 \, \right].
\end{equation}
The binary parameter $\phi_{i,j}$ represents the stable vote for the configuration. We compute the final aggregated text voting profile $v$ via double summation
\begin{equation}
v = \sum_{i=1}^{|\mathcal{C}|} \sum_{j=1}^{|\mathcal{S}|} \phi_{i,j} \;\in\; \{0, 1, \dots, 15\}.
\end{equation}
This voting profile characterises the overarching discrete pragmatic confidence.

The framework applies the Stage~1 decision rule of
Eq.~\ref{case} when acoustic streams are absent. The
nested majority voting protocol effectively smooths local token
sensitivity. This non-sequential aggregation strategy avoids heavy analytical scaffolds. It counteracts the systematic positivity bias documented in prior LLM-based sarcasm evaluations.

\subsection{Acoustic Late-Fusion Rescue (ALFR)}
\label{sec:alfr_design}
We introduce the Acoustic Late-Fusion Rescue module to recover classification performance when text alone saturates. Operating outside the black-box language network boundary, the architecture decouples textual voting properties from structural acoustic observations, which avoids the cross-modal single-class collapses that characterise conventional fusion. The module applies a downstream fusion pattern over an asynchronous feature space.

\subsubsection*{Multi-Modal Heterogeneous Feature Representation Space}
The ALFR module constructs a unified heterogeneous feature space that aggregates the linguistic calibration vote alongside raw prosody, spanning twenty-seven multimodal dimensions in total.

The first dimension encapsulates the discrete linguistic confidence metrics from Section~\ref{sec:bical_design}. Let $v \in \{0, 1, \dots, 15\}$ represent the scalar text voting profile.

The subsequent six dimensions encompass low-level physical acoustic functionals from openSMILE. Let $a \in \mathbb{R}^6$ denote the raw acoustic vector. These elements represent mean pitch, pitch variation, loudness, spectral flux, harmonic clarity, and tempo. To eliminate speaker variance without utilising labels, we implement a cascading baseline pool. Let $s$ denote the unique speaker identity. Let $d$ represent the specific dialogue context string. We determine the operative reference pool using a hierarchical validation cascade
\begin{equation}
\Omega = \begin{cases} \Omega_{\text{L1}}(s, d) & \text{if } |\Omega_{\text{L1}}(s, d)| \;\ge\; n_{\min}, \\ \Omega_{\text{L2}}(s) & \text{else if } |\Omega_{\text{L2}}(s)| \;\ge\; n_{\min}, \\ \Omega_{\text{L3}} & \text{otherwise}, \end{cases}
\end{equation}
where $\Omega_{\text{L1}}$ restricts the context to the same speaker within the dialogue. The secondary pool $\Omega_{\text{L2}}$ evaluates cross-dialogue samples for the identical speaker. The final fallback pool $\Omega_{\text{L3}}$ represents the universal baseline. The threshold parameter $n_{\min}$ equals two for the MUStARD corpus. This threshold parameter equals five for the CMMA corpus to counteract smaller per-dialogue speaker pools. We compute the normalised rolling acoustic vector $z \in \mathbb{R}^6$ feature-wise
\begin{equation}
z_m = \frac{a_m - \mu_m\bigl(\Omega \setminus \{a\}\bigr)}{\sigma_m\bigl(\Omega \setminus \{a\}\bigr) + \epsilon},
\end{equation}
where $\mu_m$ and $\sigma_m$ denote pool mean and standard deviation. The subtraction of the singleton set enforces strict causal fairness during extraction. The scaling parameter $\epsilon = 10^{-8}$ prevents mathematical division errors. To prevent divergent scores when the within-pool standard deviation approaches zero, all computed metrics are clipped to the strict interval of $[-5, +5]$.

The final twenty dimensions incorporate categorical acoustic perceptions from multi-class omni probes. Rather than forcing binary verdicts which trigger collapses, we leverage fine-grained multi-class profiling. This strategic constraint routes the audio stream into robust multi-class perception tasks. We formalise the precise implementation layout, system directives, and verbatim runtime templates in Figure~\ref{fig:omni_probes}.

\begin{figure*}[htbp]
\centering

\definecolor{probeOne}{RGB}{0,128,128}    
\definecolor{probeTwo}{RGB}{123,80,160}   
\definecolor{probeThree}{RGB}{200,110,40} 
\definecolor{probeFour}{RGB}{40,100,170}  

\newcommand{\optchip}[2]{%
  \tcbox[on line, colback=#1!12, colframe=#1, coltext=#1!50!black,
         boxsep=1pt, left=4pt, right=4pt, top=0.5pt, bottom=0.5pt,
         arc=2pt, boxrule=0.5pt, nobeforeafter]{\ttfamily\footnotesize #2}}

\newcommand{\probebadge}[2]{%
  \tcbox[on line, colback=#1, colframe=#1, coltext=white,
         boxsep=1.5pt, left=5pt, right=5pt, top=1pt, bottom=1pt,
         arc=3pt, boxrule=0pt, nobeforeafter]{\bfseries\small #2}}

\begin{tcolorbox}[enhanced, width=0.96\textwidth,
    colback=white, colframe=black!70, boxrule=0.8pt, arc=3pt,
    left=12pt, right=12pt, top=8pt, bottom=10pt,
    title={\centering\bfseries\large Multi-Class Acoustic Perception Probing Framework},
    coltitle=white, fonttitle=\bfseries, colbacktitle=black!70,
    halign title=center]

{\centering\small\itshape Omni Probes -- Runtime Configuration\par}
\vspace{4pt}

{\centering
\probebadge{probeOne}{$P_1$ : 4-way}\,\,
\probebadge{probeTwo}{$P_2$ : 8-way}\,\,
\probebadge{probeThree}{$P_3$ : 3-way}\,\,
\probebadge{probeFour}{$P_4$ : 5-pt scale}\par}
\vspace{6pt}

\begin{tcolorbox}[enhanced, colback=probeOne!4, colframe=probeOne,
    boxrule=0.5pt, arc=2pt, left=8pt, right=8pt, top=5pt, bottom=6pt,
    borderline west={3pt}{0pt}{probeOne},
    title style={fill=probeOne}, fonttitle=\bfseries, coltitle=white,
    title={$P_1$\,\,Emotional Intensity Profile \quad\ttfamily\normalfont\footnotesize P1\_intensity
           \hfill \normalfont\itshape single word, 4-way}]
  \footnotesize
  \textbf{Runtime Prompt Template:}\\
  ``Listen to this clip. How \textbf{emotionally intense} is the speaker's voice?
  Choose ONE: monotone, mild, moderate, strong. Reply with one word.''\\[3pt]
  \textbf{Options:}\,
  \optchip{probeOne}{monotone}\,\optchip{probeOne}{mild}\,%
  \optchip{probeOne}{moderate}\,\optchip{probeOne}{strong}
\end{tcolorbox}

\begin{tcolorbox}[enhanced, colback=probeTwo!4, colframe=probeTwo,
    boxrule=0.5pt, arc=2pt, left=8pt, right=8pt, top=5pt, bottom=6pt,
    borderline west={3pt}{0pt}{probeTwo},
    title style={fill=probeTwo}, fonttitle=\bfseries, coltitle=white,
    title={$P_2$\,\,Vocal Style Profiling \quad\ttfamily\normalfont\footnotesize P2\_style
           \hfill \normalfont\itshape single word, 8-way}]
  \footnotesize
  \textbf{Runtime Prompt Template:}\\
  ``Listen to this clip. Choose the ONE word that best describes the speaker's
  \textbf{vocal style}: flat, animated, dramatic, sincere, bored, excited, hostile,
  friendly. Reply with just that one word.''\\[3pt]
  \textbf{Options:}\,
  \optchip{probeTwo}{flat}\,\optchip{probeTwo}{animated}\,%
  \optchip{probeTwo}{dramatic}\,\optchip{probeTwo}{sincere}\,%
  \optchip{probeTwo}{bored}\,\optchip{probeTwo}{excited}\,%
  \optchip{probeTwo}{hostile}\,\optchip{probeTwo}{friendly}
\end{tcolorbox}

\begin{tcolorbox}[enhanced, colback=probeThree!4, colframe=probeThree,
    boxrule=0.5pt, arc=2pt, left=8pt, right=8pt, top=5pt, bottom=6pt,
    borderline west={3pt}{0pt}{probeThree},
    title style={fill=probeThree}, fonttitle=\bfseries, coltitle=white,
    title={$P_3$\,\,Tone--Text Transcript Alignment \quad\ttfamily\normalfont\footnotesize P3\_alignment
           \hfill \normalfont\itshape dynamic, 3-way}]
  \footnotesize
  \textbf{Dynamic Runtime Function} (\texttt{p3\_prompt(utterance)})\textbf{:}\\
  ``Listen to this clip. The transcript of what is said is: ``\texttt{\{utterance\}}''.
  How well does the speaker's \textbf{vocal tone match the literal meaning} of those
  words? Choose ONE: matches, mismatched, unclear. Reply with one word.''\\[3pt]
  \textbf{Options:}\,
  \optchip{probeThree}{matches}\,\optchip{probeThree}{mismatched}\,%
  \optchip{probeThree}{unclear}
\end{tcolorbox}

\begin{tcolorbox}[enhanced, colback=probeFour!4, colframe=probeFour,
    boxrule=0.5pt, arc=2pt, left=8pt, right=8pt, top=5pt, bottom=6pt,
    borderline west={3pt}{0pt}{probeFour},
    title style={fill=probeFour}, fonttitle=\bfseries, coltitle=white,
    title={$P_4$\,\,Speaker Expressiveness Index \quad\ttfamily\normalfont\footnotesize P4\_expressiveness
           \hfill \normalfont\itshape numeric, 5-pt scale}]
  \footnotesize
  \textbf{Runtime Prompt Template:}\\
  ``Listen to this clip. Rate the speaker's \textbf{vocal expressiveness} on a scale of
  1 to 5 (1 = very monotone, 5 = highly expressive). Reply with one number.''\\[3pt]
  \textbf{Options:}\,
  \optchip{probeFour}{1}\,\optchip{probeFour}{2}\,\optchip{probeFour}{3}\,%
  \optchip{probeFour}{4}\,\optchip{probeFour}{5}
\end{tcolorbox}

\end{tcolorbox}
\caption{Verbatim runtime templates for the four Multi-Class Omni Perception Probes.
Each probe isolates the exact prompt template executed in the production script; the
title badge reports its output space (number of categorical options or scale points),
and admissible answers are listed as chips. Probe $P_3$ is a \emph{dynamic} directive
that splices the character-bounded transcript string \texttt{\{utterance\}} into the
listening prompt for cross-modal verification. Key perceptual targets are highlighted
in \textbf{bold}; dynamic input slots are set in a typewriter typeface.}
\label{fig:omni_probes}
\end{figure*}

Let $P_1$ track voice intensity across four discrete options. Let $P_2$ track vocal style across eight discrete choices. Let $P_3$ evaluate tone matching across three choices. Let $P_4$ monitor expressiveness across five numerical ratings. Let $\mathbf{o} \in \{0, 1\}^{20}$ denote the concatenated dense binary vector. We map each indicator through a standalone one-hot encoding (denoted as $oh$) function
\begin{equation}
\mathbf{o} = \left[ oh(P_1)^T, \, oh(P_2)^T, \, oh(P_3)^T, \, oh(P_4)^T \right]^T.
\end{equation}

\subsubsection*{Feature Assembly and Classification Routing}
We construct the final joint vector by concatenating these heterogeneous input matrices. Let $x_{\text{joint}} \in \mathbb{R}^{27}$ denote the integrated feature instance
\begin{equation}
x_{\text{joint}} = \left[ \begin{array}{c} v \\ z \\ \mathbf{o} \end{array} \right].
\end{equation}
This concatenation preserves the interpretability of each individual channel while jointly exposing linguistic and acoustic cues to the classifier.

The combined multimodal vector feeds a parameterised shallow meta-learner that optimises the secondary classification boundary. Let $\mathcal{M}$ represent the generalised function mapping this vector to the label space
\begin{equation}
\hat{y}^{(2)} = \mathcal{M}\bigl(x_{\text{joint}}\bigr) \in \{0, 1\}.
\end{equation}
Implemented as either a linearly regularised model or a non-linear boosting ensemble, $\mathcal{M}$\ accommodates varying target priors across language environments and adaptively re-weights features. This flexibility lets the framework rescue over-firing backbones by downweighting saturated language-model votes in favour of the orthogonal acoustic features. 
\section{Experimental Setup}
\subsection{Datasets and Evaluation Profile}
\label{sec:datasets_profile}
We evaluate our framework on two multimodal sarcasm benchmarks across different linguistic environments. The evaluation utilises the English MUStARD \cite{castro2019towards} corpus alongside the Chinese CMMA \cite{zhang2023cmma} dataset. We formalise the statistical profiles of these evaluation corpora in Table~\ref{tab:dataset_stats}.

\begin{table}[htbp]
\caption{Statistical Profiles of the Evaluation Datasets}
\label{tab:dataset_stats}
\centering
\small
\begin{tabular}{lccc}
\hline
\textbf{Dataset} & \textbf{Sample Size} & \textbf{Sarcasm Ratio} & \textbf{Language} \\ \hline
MUStARD & 690 & 50.0\% & English \\
CMMA & 3961 & 11.4\% & Chinese \\ \hline
\end{tabular}
\end{table}

MUStARD comprises 690 conversational video clips drawn from popular English-language sitcoms. These series include \textit{Friends}, \textit{The Big Bang Theory}, \textit{The Golden Girls}, and \textit{Sarcasmaholics Anonymous}. Owing to the limited size of MUStARD, the literature lacks a single standardised train–test partition \cite{castro2019towards, gao2024amused, saini2024grainy}. Since our text-only calibration stage is entirely training-free, we evaluate this phase on the full set of 690 clips. In contrast, the lightly-supervised ALFR stage is assessed via stratified 5-fold cross-validation. 

CMMA is a large-scale Chinese multi-party conversational benchmark. This corpus aggregates dialogues from popular domestic series such as \textit{iPartment}, \textit{Three Kingdoms}, and \textit{Cooking Class Story}. The dataset is severely class-imbalanced, with sarcastic utterances comprising only 11.4\% of the cases. We adopt the official test split of CMMA, comprising 3,961 utterances, rather than the substantially larger full corpus. This choice keeps our results directly comparable with the standard evaluation protocol of the benchmark.

We enforce a unified partitioning protocol across experiments. The standalone text calibration phase processes the full un-split dataset and consumes no downstream task labels. Conversely, the late-fusion phase runs a stratified cross-validation scheme. The acoustic extractors and multi-class perceptual probes operate strictly within these designated evaluation subsets.

\subsection{Models and Framework Components}
\label{sec:models_framework}
The experimental framework aggregates foundational language networks, acoustic functional extractors, and downstream regularised routing heads. This modular setup evaluates performance across varying parameter scales and architecture designs.

\subsubsection*{Linguistic Foundation Backbones}
We employ seven LLM backbones to validate text-based performance, which comprise Qwen2.5-32B-Instruct \cite{qwen2025qwen25technicalreport},  Qwen2.5-7B-Instruct \cite{qwen2025qwen25technicalreport}, Llama-3.1-8B-Instruct \cite{grattafiori2024llama}, Gemma-2-27B-it \cite{team2024gemma}, Mistral-Small-Instruct-2409 \cite{jiang2024mixtral}, DeepSeek-V4-Pro \cite{liu2024deepseek}, and GPT-5.4 \cite{singh2025openai}. All linguistic backbones process input prompts under identical prompting configurations to maintain evaluation parity.

\subsubsection*{Acoustic Modality Components}
The acoustic processing stream decouples speech signals into complementary physical and perceptual feature spaces. Low-level physical prosody functionals are captured through the openSMILE toolkit \cite{eyben2010opensmile}, which executes the standardised eGeMAPSv02 configuration pool. High-level categorical acoustic perceptions are extracted using the Qwen2.5-Omni-7B \cite{xu2025qwen2} multimodal architecture acting as zero-shot omni probes. The framework isolates these frozen feature extractors to prevent foundational parameter updates during downstream evaluation.

\subsubsection*{Lightweight Decision Classifiers}
The parameterised decision routing block outside the foundational network boundary aggregates a candidate pool of ten classic classification models. This evaluation pool is partitioned into eight individual estimators and two meta-ensemble architectures. The individual estimators encompass Logistic Regression (LR) \cite{hosmer2013applied}, Random Forests (RF) \cite{breiman2001random}, Gradient Boosting (GB) \cite{friedman2001greedy}, and AdaBoost \cite{freund1997decision}. This group also includes Support Vector Machines (SVM) \cite{drucker1996support}, Multi-Layer Perceptrons (MLP) \cite{popescu2009multilayer}, XGBoost \cite{chen2016xgboost}, and LightGBM \cite{ke2017lightgbm}. The meta-ensemble configurations incorporate Stacking protocols (Stacking) \cite{wolpert1992stacked} and Heterogeneous Soft Voting \cite{kittler1998combining}. The parameter structures of all ten estimators remain fixed to standardised library configurations to prevent over-parameterised fitting leakage, and the meta-learners resolve decision boundaries solely within the cross-validation training partitions. We relegate the precise parameter specifications of all ten candidate classifiers to Appendix~\ref{sec:classifier_hyperparameters}.

\subsection{Implementation Details}
\label{sec:implementation_details}
This section formalises the precise operational parameters across our execution pipeline. We consolidate the parameter layout to ensure experimental transparency and full reproducibility.

\subsubsection{Large Language Model Inference Configuration}
\label{sec:llm_inference_config}
The experimental framework establishes a unified sampling profile across all evaluated language networks. We adopt Direct Zero-Shot Prompting (DZS) \cite{brown2020language} as the standard reference configuration. This protocol matches the baseline architecture used in recent LLM-based sarcasm evaluation \cite{mai2024llama}. All subsequent BiCAL and ALFR improvements are reported directly against this reference. DZS employs a single prompt template executed via greedy decoding by setting $\mathtt{do\_sample}$ to false, allocating one sample per utterance. All stochastic generation, by contrast, follows a standardised sampling layout: temperature 0.5, top-p 0.95, three response samples per prompt-seed combination, and three replication seeds (42, 43, 44), with final predictions resolved by majority voting. Generation is capped at twelve tokens, and a deterministic parser extracts the binary classification indicator.

\subsubsection{Acoustic Feature Extraction Pipelines}
\label{sec:acoustic_feature_pipelines}
The acoustic processing stream executes the dual extraction channels formalised in Section 3. For the physical channel, the openSMILE software engine extracts the 6-dimensional eGeMAPSv02 baseline functional vector. The hierarchical cascade pool $\Omega$ applies the dataset-specific threshold parameter $n_{\min}$ to calculate un-biased z-scores. For the perceptual channel, the frozen Qwen2.5-Omni-7B multimodal network acts as zero-shot probes. The omni engine executes strict greedy decoding to generate one single deterministic response per diagnostic query. Standalone one-hot encoding mapping tables transform these categorical outputs into a 20-dimensional perception vector. 

\subsubsection{Classifier Optimisation and Layered Supervision}
\label{sec:classifier_optimization_supervision}
The framework operates under a strictly layered supervision profile to separate representations from routing. The linguistic calibration blocks and acoustic extraction modules remain entirely zero-shot. The post-network late fusion layer consumes downstream task labels exclusively. This regularised learning layer maps the 27-dimensional concatenated multimodal feature vector to the target label space. We configure a standard AdaBoost classifier for the symmetric English MUStARD dataset and deploy a Heterogeneous Soft Voting ensemble for the Chinese CMMA benchmark. The downstream training pipeline runs a five-fold stratified cross-validation sequence tracking five independent random seeds from 42 to 46. All candidate classifiers maintain fixed standard factory parameter settings to avoid optimisation bias.

\subsubsection{Statistical Testing Framework}
\label{sec:statistical_testing_framework}
We employ a paired bootstrap resampling protocol to evaluate performance variations reliably. The statistical engine executes 10,000 bootstrap replication cycles, resampling unique sample identifiers instead of cross-validation folds. The evaluation computes the Stouffer Z meta-analysis metric \cite{whitlock2005combining} across the seven LLM backbones, with the Bonferroni correction \cite{armstrong2014use} applied to handle family-wise error rates.

All implementation components are fully documented to support reproducibility. The random seeds, foundational model identifiers, and dataset partitions are public. The complete production source code repository is hosted online\footnote{\url{https://github.com/glam-imperial/charm}}. We compile comprehensive model cards in Appendix~\ref{app:model_cards}.

\section{Main Results and Performance Analysis}
\label{sec:main_results}
This section evaluates CHARM across diverse linguistic environments. We first assess the standalone performance of the linguistic BiCAL layer under zero-shot textual constraints, followed by a systematic evaluation of the ALFR module against direct zero-shot baselines. Table~\ref{tab:main_results} reports the consolidated results across two datasets.

\begin{table*}[t]
\centering
\caption{
Stage-wise performance progression of CHARM on MUStARD and CMMA benchmarks reported using Macro-F1 and Acc. metrics alongside the optimal late-fusion routing head. DZS denotes direct zero-shot prompting with greedy decoding. This configuration serves as the reference baseline from prior LLM-based sarcasm evaluations~\cite{11146812,mai2024llama}. BiCAL applies bidirectional linguistic calibration. ALFR denotes the proposed acoustic late-fusion rescue module. Best performance within each backbone family is highlighted in bold. Stouffer $Z$ statistic aggregates per-model paired-bootstrap z-scores across all seven backbones.
}
\label{tab:main_results}
\small
\setlength{\tabcolsep}{10pt}
\renewcommand{\arraystretch}{1.15}
\begin{tabular}{lccccccc}
\toprule
& \multicolumn{2}{c}{DZS} & \multicolumn{2}{c}{+BiCAL} & \multicolumn{3}{c}{+ALFR} \\
\cmidrule(lr){2-3}\cmidrule(lr){4-5}\cmidrule(lr){6-8}
LLM Backbone Family & Macro-F1 & Acc. & Macro-F1 & Acc. & Macro-F1 & Acc. & Best CLF \\
\midrule
\multicolumn{8}{l}{\textit{\textbf{MUStARD}}} \\
\midrule
Mistral-Small-Instruct-2409 & 0.718 & 0.723 & \textbf{0.736} & \textbf{0.736} & 0.729 & 0.730 & LR \\
Qwen2.5-32B-Instruct       & 0.699 & 0.700 & 0.713 & 0.713 & \textbf{0.729} & \textbf{0.730} & LR \\
Gemma-2-27B-it        & 0.447 & 0.549 & 0.470 & 0.562 & \textbf{0.721} & \textbf{0.725} & AdaBoost \\
Qwen2.5-7B        & 0.409 & 0.519 & 0.420 & 0.523 & \textbf{0.635} & \textbf{0.636} & RF \\
Llama-3.1-8B-Instruct      & 0.401 & 0.529 & 0.411 & 0.533 & \textbf{0.635} & \textbf{0.641} & LGBM \\
DeepSeek-V4-Pro   & 0.742 & 0.743 & \textbf{0.787} & \textbf{0.787} & 0.784 & 0.785 & Soft Voting \\
GPT-5.4           & 0.654 & 0.678 & 0.730 & 0.736 & \textbf{0.787} & \textbf{0.787} & Stacking \\
\midrule
\multicolumn{1}{l}{Stouffer $Z$ (DZS $\rightarrow$ +BiCAL)} & \multicolumn{7}{c}{$Z = 5.85,\ p = 2.4 \times 10^{-9}$} \\
\multicolumn{1}{l}{Stouffer $Z$ (+BiCAL $\rightarrow$ +ALFR)} & \multicolumn{7}{c}{$Z = 13.89,\ p < 10^{-43}$} \\
\midrule
\multicolumn{8}{l}{\textit{\textbf{CMMA}}} \\
\midrule
Mistral-Small-Instruct-2409 & 0.527 & 0.637 & 0.572 & 0.723 & \textbf{0.591} & \textbf{0.823} & Soft Voting \\
Qwen2.5-32B-Instruct       & 0.577 & 0.740 & \textbf{0.599} & 0.796 & 0.594 & \textbf{0.824} & Soft Voting \\
Gemma-2-27B-it        & 0.273 & 0.275 & 0.286 & 0.290 & \textbf{0.607} & \textbf{0.840} & Soft Voting \\
Qwen2.5-7B        & 0.269 & 0.271 & 0.279 & 0.282 & \textbf{0.586} & \textbf{0.836} & Soft Voting \\
Llama-3.1-8B-Instruct      & 0.144 & 0.149 & 0.193 & 0.194 & \textbf{0.575} & \textbf{0.831} & Soft Voting \\
DeepSeek-V4-Pro   & 0.542 & 0.664 & 0.592 & 0.758 & \textbf{0.619} & \textbf{0.829} & Soft Voting \\
GPT-5.4           & 0.545 & 0.647 & 0.604 & 0.768 & \textbf{0.627} & \textbf{0.819} & Soft Voting \\
\midrule
\multicolumn{1}{l}{Stouffer $Z$ (DZS $\rightarrow$ +BiCAL)} & \multicolumn{7}{c}{$Z = 19.29,\ p < 10^{-50}$} \\
\multicolumn{1}{l}{Stouffer $Z$ (+BiCAL $\rightarrow$ +ALFR)} & \multicolumn{7}{c}{$Z = 34.64,\ p < 10^{-50}$} \\
\bottomrule
\end{tabular}
\end{table*}

\subsection{Stage 1 Evaluation: Standalone Linguistic Calibration Performance}
\label{sec:stage1_evaluation}
The first stage assesses the standalone bidirectional charge-calibration layer under zero-shot textual constraints. Table~\ref{tab:main_results} shows a uniform positive progression: BiCAL produces positive Macro-F1 increments for all seven backbones on both datasets, consistently across structurally distinct model families and parameter scales. Consistent with a calibration mechanism, the size of each gain tracks the backbone's intrinsic over-prediction. The heavily over-firing GPT-5.4 backbone, which exhibits a positive class prediction rate of 76.7\%, benefits most. Its Macro-F1 score improves by an absolute margin of 0.076 on MUStARD ($0.654 \to 0.730$). Conversely, better-aligned architectures improve more modestly. The Macro-F1 score of Gemma-2-27B-it increases by 0.023 ($0.447 \to 0.470$), while DeepSeek-V4-Pro yields a Macro-F1 gain of 0.045 ($0.742 \to 0.787$). In absolute terms, the calibrated DeepSeek-V4-Pro sets the overall zero-shot text-only peak at a Macro-F1 score of 0.787, followed by Mistral-Small-Instruct-2409 at 0.736.

This calibration trend generalises across highly divergent linguistic environments. Applying the English templates verbatim to the CMMA keeps the Macro-F1 gains consistently positive across all seven backbones, ranging from $+0.010$ for Qwen2.5-7B ($0.269 \to 0.279$) to $+0.059$ for GPT-5.4 ($0.545 \to 0.604$). This stable cross-lingual transferability validates the bidirectional formulation, demonstrating that the charge structure neutralises the instruction-following bias regardless of the underlying tokenisation language.

We verify these gains with a paired bootstrap resampling protocol of 10,000 cycles, deriving per-backbone z-scores (observed delta over standard error) across all seven backbones.
A cumulative Stouffer meta-analysis of these z-scores proves highly significant. The analysis yields $Z = 5.85$ with $p = 2.4\times 10^{-9}$ on MUStARD, which rises to $Z = 19.29$ with $p < 10^{-50}$ on CMMA. The more than threefold larger $Z$ on CMMA reflects the substantially wider calibration headroom available under extreme positive-class imbalance.

The joint behaviour of Macro-F1 and Accuracy is also informative. On the balanced MUStARD, the two metrics stay tightly coupled within a 0.005 interval for every model family, indicating that the lift reflects genuine probability calibration rather than an exploitation of distributional skew. On the highly imbalanced CMMA benchmark, the larger gap between Macro-F1 and Accuracy is a natural consequence of the 11.4\% sparse positive prior and does not compromise the validity of the calibrated Macro-F1 progression.

The text-only calibration gain is structurally bounded between $+0.010$ and $+0.076$ Macro-F1 across all backbones, with the largest lift accruing to the most heavily over-predicting model and the smallest to the already-calibrated ones. This bounded, over-prediction-tracking lift matches our design intent, confirming that BiCAL acts purely as a statistical calibration layer that removes RLHF sycophancy, without adding external features or altering the frozen model's reasoning capacity. The capability gap between strong and weak text models therefore persists in Stage 1. On CMMA, the Macro-F1 gap between Llama-3.1-8B-Instruct at 0.193 and GPT-5.4 at 0.604 remains a wide 0.411. This closely mirrors the 0.401 gap the two models exhibit in the uncalibrated DZS phase, where they score 0.144 and 0.545, respectively. Bidirectional calibration is thus necessary but insufficient for text-dominated classifiers. Small models lack the zero-shot reasoning capacity to resolve implicit cross-cultural irony from text alone, and this textual ceiling motivates the Stage 2 acoustic late-fusion rescue.

\subsection{Stage 2 Evaluation: Multi-Modal Late-Fusion Rescue}
\label{sec:stage2_evaluation}
Integrating acoustic features produces a pronounced dichotomy across backbones, proving that the late-fusion module acts as a rescue mechanism rather than a generic retrofit. On MUStARD, Gemma-2-27B-it achieves an absolute Macro-F1 improvement of 0.251 to reach 0.721. Meanwhile, Llama-3.1-8B-Instruct and Qwen2.5-7B exhibit Macro-F1 increases of 0.224 and 0.215 relative to their respective Stage 1 scores. The effect intensifies on the imbalanced CMMA benchmark. Here, Llama-3.1-8B-Instruct records a peak single-model absolute Macro-F1 increase of 0.382 to reach a score of 0.575, with Gemma-2-27B-it and Qwen2.5-7B following with gains of 0.321 and 0.307, respectively.

Mistral-Small-Instruct-2409 and Qwen2.5-32B-Instruct are already saturated, varying by under 0.020 in Macro-F1 score across either environment. DeepSeek-V4-Pro is likewise saturated on MUStARD and remains essentially unchanged. Conversely, the mildly over-predicting GPT-5.4 still retains acoustic headroom and is rescued by an absolute Macro-F1 increment of 0.057 to reach a score of 0.787 on MUStARD. On the imbalanced CMMA benchmark, both GPT-5.4 and DeepSeek-V4-Pro achieve a modest absolute increase of 0.023 in Macro-F1 score. A cumulative Stouffer meta-analysis across all seven backbones confirms the significance of this rescue phase. The analysis yields $Z = 13.89$ with $p < 10^{-43}$ on MUStARD, scaling to $Z = 34.64$ with $p < 10^{-50}$ on CMMA. The larger Chinese dataset reflects the expanded rescue space available for heavily collapsed baselines.

This multimodal progression also reveals a convergence effect across model families. Computing the Pearson correlation between the zero-shot text baseline and the total Macro-F1 increment delivered by the full pipeline across all seven backbones uncovers a strong linear inversion: $r = -0.946$ on MUStARD and $r = -0.996$ on CMMA. Backbones with the weakest text baselines receive the largest acoustic rescue, while already-strong backbones move little; GPT-5.4 retains a positive increment owing to its residual over-prediction, whereas the saturated DeepSeek-V4-Pro sits near the zero-increment end of the line.

The optimal fusion routing head is data-driven and varies with the dataset's class distribution. On the MUStARD dataset, the choice is heterogeneous and selected per backbone: Mistral-Small-Instruct-2409 and Qwen2.5-32B-Instruct peak with standard Logistic Regression, whereas the remaining five backbones favour non-linear ensemble heads, such as AdaBoost, Random Forests, and LightGBM for Gemma-2-27B-it, Qwen2.5-7B, and Llama-3.1-8B, and Stacking and soft voting for GPT-5.4 and DeepSeek-V4-Pro, respectively. The diversity of optimal heads underscores that the routing is learnt rather than fixed.

On the imbalanced CMMA corpus, the fusion layer converges onto a homogeneous profile. The heterogeneous soft voting ensemble dominates across all text networks, outperforming the second-best standalone Random Forest by a consistent margin of 0.017 to 0.045 in Macro-F1 score. Standard boosting architectures collapse into the majority negative class under severe sparsity; the standalone AdaBoost model, for instance, matches the dummy baseline exactly. It scores a high Accuracy of 0.886 with a degraded Macro-F1 of 0.470. Soft voting overcomes this by integrating three distinct inductive biases together with balanced class penalisation. We compile the complete optimisation sweeps for all ten candidate fusion
classifiers across backbones in Appendix~\ref{app:classifier_sweep}.

We note the contrasting trajectories of Accuracy and Macro-F1 on the imbalanced Chinese benchmark. The framework's Accuracy across the seven backbones stays between 0.819 and 0.840, slightly below the always-negative majority baseline of 0.886. This is a deliberate trade-off rather than a regression, as the balanced class weighting penalises majority-class errors to maximise minority sarcastic recall. A naive majority-class baseline reaches an Accuracy score of 0.886 but a stagnant Macro-F1 score of only 0.470. The late-fusion rescue breaks this passive saturation, delivering absolute Macro-F1 elevations of 0.105 to 0.157 over that baseline across all seven backbones. Trading marginal majority-class accuracy for robust sensitivity to rare sarcastic instances aligns with established practice for skewed distributions, where Macro-F1 is the primary efficacy metric.

\subsection{Comparison with Prior Work}
\label{sec:sota}
\begin{table}
\centering
\caption{Comparison with prior sarcasm-detection methods on MUStARD and CMMA.
Modality: T(ext), A(udio), V(ideo). Training: \emph{Sup.}\ (supervised on the
benchmark), \emph{SFT} (LLM Supervised fine-tuning), \emph{Train-free} (zero-shot prompting),
\emph{Light-sup.}\ (frozen backbones with a shallow trained head). ``--'' denotes a
metric not reported by the source. Our CMMA scores use the English-prompt setting
on the audio-available subset; MUStARD follows the original split. The best result
among training-free methods is in bold.
}
\label{tab:sota}
\small
\setlength{\tabcolsep}{8pt}
\renewcommand{\arraystretch}{1.15}
\resizebox{\linewidth}{!}{%
\begin{tabular}{l c c c c c}
\toprule
Method & Modality & Training & Macro-F1 & Acc & F1 \\
\midrule
\multicolumn{6}{l}{\textit{MUStARD}} \\
\midrule
Sukhavasi et al.~\cite{sukhavasi2025deep} & T, A, V & Sup.        & --    & --    & 0.730 \\
Xue et al.~\cite{xue2024breakthrough}      & T, A, V & Sup.        & --    & 0.769  & 0.761 \\
Buaroiu et al.~\cite{buaroiu2023capable}   & T     & SFT         & --    & --    & 0.770 \\
Yao et al.~\cite{yao2025sarcasm}           & T     & Train-free  & 0.699  & 0.707  & --   \\
Lee et al.~\cite{lee2025pragmatic}         & T     & Train-free  & 0.777  & \textbf{0.794}  & --   \\
Zheng et al.~\cite{zheng2025masd}          & T     & Train-free  & 0.762  & 0.762  & --   \\
Xiong et al.~\cite{xiong2025sarc7}         & T     & Train-free  & --    & --    & 0.366 \\
\textbf{CHARM-BiCAL (Ours)}                & T     & Train-free  & \textbf{0.787} & 0.787 & --   \\
\textbf{CHARM-ALFR (Ours)}                 & T, A   & Light-sup.  & 0.784  & 0.785  & --   \\
\midrule
\multicolumn{6}{l}{\textit{CMMA}} \\
\midrule
Zhang et al.~\cite{zhang2023cmma}          & T     & Sup.        & 0.541 & --    & --   \\
Zhang et al.~\cite{zhang2023cmma}          & T, A   & Sup.        & 0.532 & --    & --   \\
Zhang et al.~\cite{zhang2023cmma}          & T, A, V & Sup.        & 0.752 & --    & --   \\
\textbf{CHARM-BiCAL (Ours)}                & T     & Train-free  & \textbf{0.604} & 0.768 & --   \\
\textbf{CHARM-ALFR (Ours)}                 & T, A   & Light-sup.  & 0.627  & 0.819 & --   \\
\bottomrule
\end{tabular}%
}
\end{table}

Table~\ref{tab:sota} situates \textsc{CHARM} among prior sarcasm detection methods. This comparison is necessarily heterogeneous, since the listed systems differ in supervision regime, modality, and evaluation protocol; the table therefore offers broader context rather than a like-for-like leaderboard, and we evaluate \textsc{CHARM} primarily against methods that share its validation setting.

On MUStARD, \textsc{CHARM}-BiCAL attains the highest Macro-F1 score of 0.787 among training-free approaches. This performance surpasses the metacognitive prompting of Lee et al.~\cite{lee2025pragmatic}, which yields a Macro-F1 score of 0.777, as well as the step-by-step reasoning of Yao et al.~\cite{yao2025sarcasm}, which achieves a Macro-F1 score of 0.699. It achieves this through a lightweight bidirectional calibration rather than an elaborate multi-step reasoning scaffold and without any fine-tuning. The resulting score is competitive even with fully supervised multimodal systems that additionally exploit the visual stream~\cite{xue2024breakthrough,sukhavasi2025deep}, despite relying on text alone.

On CMMA, no prior work evaluates zero-shot LLMs, so we compare against the supervised baselines of the original benchmark~\cite{zhang2023cmma}, which require task-specific training. In this scenario, \textsc{CHARM}-BiCAL achieves a Macro-F1 score of 0.604, which already exceeds the supervised text-only baseline score of 0.541 and the supervised audio-text baseline score of 0.532. Furthermore, \textsc{CHARM}+ALFR widens this margin to reach a peak Macro-F1 score of 0.627.

Tellingly, the fully supervised baseline derives no benefit from acoustics. Its audio-text score of 0.532 falls below its text-only performance score of 0.541. In contrast, ALFR secures a consistent gain from 0.604 to 0.627, directly corroborating our late-fusion design. Only the fully supervised tri-modal configuration of Zhang et al. \cite{zhang2023cmma} 
remains ahead, achieving a superior Macro-F1 score of 0.752 by exploiting visual cues. Visual streams introduce rich pragmatic features, such as facial micro-expressions and body postures, which heavily facilitate supervised intent recognition. However, processing full-scale video tracking significantly escalates computational complexity and parametric overhead. We intentionally exclude the visual modality to focus exclusively on resolving the linguistic-prosodic interface within frozen foundation models, allowing CHARM to maintain a streamlined and lightweight architecture while bypassing the spatial noise inherent to video tracking pipelines.

These outcomes require two distinct caveats. Our CMMA evaluation uses the audio-available subset under the English-prompt setting. Additionally, MUStARD scores reported across prior literature vary based on the specific data split and corpus variant.

\subsection{Ablation Studies}
\label{sec:ablation_studies}
Three targeted ablation experiments substantiate the core structural choices of CHARM. We first demonstrate that the performance gain of BiCAL stems from charge polarity rather than from prompt diversity. The evaluation then confirms that the three distinct feature blocks of ALFR are individually insufficient yet jointly necessary. Finally, we characterise the cost-accuracy frontier of the charge ensemble to identify two valid operating points that balance inference overhead against stochastic variance protection.

\subsubsection{Random Paraphrase Control}
\label{sec:random_paraphrase_control}
We isolate whether the zero-shot textual lift of BiCAL over the direct baseline derives from charge polarity or from generic prompt and sampling diversity. Under a fixed sampling profile matching our framework parameters, we replace the five charged templates with five neutral paraphrases compiled in Appendix~\ref{app:randpara}. This control baseline replicates the fifteen validation votes using three random seeds on the Qwen2.5-32B-Instruct backbone.

\begin{table}[h]
\centering
\caption{Linguistic configuration ablation on the MUStARD benchmark using Qwen2.5-32B-Instruct.}
\label{tab:ablation_randpara}
\small
\setlength{\tabcolsep}{14pt}
\renewcommand{\arraystretch}{1.15}
\resizebox{\columnwidth}{!}{%
\begin{tabular}{lcc}
\toprule
Linguistic Configuration Profile & Vote Count & Macro-F1 \\
\midrule
DZS Baseline & 1 & 0.699 \\
E\_maj15-RandPara Control & 15 & 0.692 \\
E\_maj15-BiCAL Framework & 15 & \textbf{0.713} \\
\bottomrule
\end{tabular}%
}
\end{table}

As documented in Table~\ref{tab:ablation_randpara}, the neutral paraphrase control yields a Macro-F1 score of 0.692. This performance remains statistically indistinguishable from the single-prompt baseline score of 0.699. The random paraphrase control falls 0.021 below the BiCAL framework score of 0.713. Paired bootstrap resampling confirms this performance deficit is statistically significant at $p < 0.05$. The observed zero-shot textual lift is therefore directly attributable to the systematic polarity of our charge configuration rather than generic prompt variety or sampling smoothing.

\subsubsection{ALFR Feature Component Ablation}

\label{sec:feature_component_ablation}
We evaluate the separate contributions of the three multimodal feature blocks within the ALFR routing layer. The evaluation isolates performance across seven additive feature subsets using the optimal downstream classifier heads evaluated across five validation seeds.
\begin{table}[h]
\centering
\caption{Ablation of the twenty-seven multimodal feature components across benchmarks using Qwen2.5-32B-Instruct. Each subset is evaluated through the optimal late-fusion head. For CMMA, this ablation uses the Chinese-prompt condition, under which the zero-shot baseline is more severely miscalibrated, leaving more headroom to probe the calibration effect.}
\label{tab:ablation_features}
\large 
\setlength{\tabcolsep}{5pt}
\renewcommand{\arraystretch}{1.15}
\resizebox{\columnwidth}{!}{%
\begin{tabular}{lcc}
\toprule
Configuration Subset & MUStARD Macro-F1 & CMMA Macro-F1 \\
\midrule
Text (Vote) & 0.713 & 0.606 \\
Acoustic & 0.598 & 0.472 \\
Perceptual & 0.569 & 0.542 \\
Text $+$ Acoustic & 0.722 & 0.615 \\
Text $+$ Perceptual & 0.718 & 0.618 \\
Acoustic $+$ Perceptual & 0.619 & 0.547 \\
Text $+$ Acoustic $+$ Perceptual (ALFR) & \textbf{0.729} & \textbf{0.622} \\
\bottomrule
\end{tabular}%
}
\end{table}

The empirical patterns recorded in Table~\ref{tab:ablation_features} demonstrate that each feature block is individually insufficient yet highly additive when combined. Every standalone acoustic or perceptual block remains bounded below a strict Macro-F1 ceiling of 0.610 across both benchmarks. Conversely, the full integration vector achieves peak Macro-F1 scores of 0.729 on MUStARD and 0.622 on CMMA, exceeding any pairwise subset variant. The performance asymmetry between linguistic environments is informative. On the MUStARD corpus, the combination of text and low-level prosody edges out the perception-probe subset by an absolute margin of 0.004 in Macro-F1 score, suggesting high diagnostic value for raw acoustics in English sitcom speech. On the CMMA, this ordering reverses, as the perception-probe combination outperforms the prosody variant by an absolute margin of 0.003 in Macro-F1 score, highlighting a cross-cultural decoupling effect.

\subsubsection{Charge Ensemble Size and Cost Analysis}
\label{sec:charge_ensemble_size}
We characterise the cost-accuracy frontier of the bidirectional charge ensemble on the Qwen2.5-32B-Instruct architecture. The ensemble voter cap $N$ is systematically varied across odd integers from 1 to 15 to prevent voting deadlocks.

\begin{table}[h]
\centering
\caption{Cost and accuracy frontier sweeps across different charge ensemble scales using Qwen2.5-32B-Instruct. Best result per benchmark in bold. $^{\dagger}$\,denotes a configuration statistically indistinguishable from the E\_maj15 reference on \emph{both} benchmarks (two-sided paired bootstrap, $p=0.26$ on MUStARD and $p=0.74$ on CMMA); the narrower three-charge E\_maj9 subset falls significantly below E\_maj15 ($p=0.002$ and $p<0.001$). For CMMA, this ablation again uses the Chinese-prompt condition (cf.\ Table~\ref{tab:ablation_features}).
}
\label{tab:ablation_ensemble_size}
\large
\setlength{\tabcolsep}{3pt}
\renewcommand{\arraystretch}{1.15}
\resizebox{\columnwidth}{!}{%
\begin{tabular}{llcc}
\toprule
Config & Operational Setup Profile & MUStARD Macro-F1 & CMMA Macro-F1 \\
\midrule
Baseline & DZS 1 Charge with Greedy Decoding & 0.699 & 0.497 \\
E\_maj3  & 1 Charge tracked across 3 Seeds   & 0.701 & 0.499 \\
E\_maj5  & 5 Charges tracked across 1 Seed   & 0.709$^{\dagger}$ & \textbf{0.598}$^{\dagger}$ \\
E\_maj9  & 3 Charges tracked across 3 Seeds   & 0.699 & 0.527 \\
E\_maj15 & 5 Charges tracked across 3 Seeds   & \textbf{0.713} & \textbf{0.598} \\
\bottomrule
\end{tabular}%
}
\end{table}

The empirical sweeps in Table~\ref{tab:ablation_ensemble_size} reveal three key insights. First, restricting the ensemble to a narrower three-charge subset (E\_maj9) significantly underperforms the full-axis E\_maj15 reference on both benchmarks. On MUStARD, the Macro-F1 score drops from 0.713 to 0.699, yielding a two-sided paired bootstrap significance level of $p = 0.002$. On CMMA, the Macro-F1 score decreases from 0.598 to 0.527 at $p < 0.001$. This trend demonstrates that charge breadth covering the full symmetric axis drives the calibration gain rather than mere seed-level sample replication.

Second, the compact E\_maj5 variant tracks five charges across a single seed. This configuration is statistically indistinguishable from E\_maj15 regarding accuracy. The performance difference is minimal on MUStARD at an absolute Macro-F1 discrepancy of -0.004 with $p = 0.260$. The variation is entirely negligible on CMMA with $p = 0.740$. Crucially, the E\_maj5 setup requires only one-third of the baseline inference budget.

Third, these two equivalent-accuracy operating points serve distinct deployment purposes. We retain E\_maj15 as the standard reference configuration throughout this paper. Its three-seed averaging minimises single-run stochastic variance. This variance reduction serves as a prerequisite for the seed-level bootstrap and Stouffer meta-analyses underpinning our significance claims. Conversely, E\_maj5 is the recommended configuration for cost-bound deployments where formal statistical inference is not required.

\subsection{Cross-cultural Analysis}
\label{sec:cross_cultural_analysis}
This section evaluates the cross-cultural generalisability of CHARM beyond English language environments through acoustic feature decoupling and prompt-language transferability tests.

\subsubsection{Acoustic Feature Decoupling}
\label{sec:acoustic_feature_decoupling}
We evaluate the cross-cultural transferability of the acoustic features by computing the effect sizes via Cohen $d$ metrics and percentage point changes across the MUStARD and CMMA environments. This analysis contrasts low-level prosodic functionals against high-level auditory language model perception probes.

\begin{table}[h]
\centering
\caption{Cross-cultural effect sizes of low-level openSMILE prosodic features.}
\label{tab:cross_cultural_low_level}
\large
\setlength{\tabcolsep}{3pt}
\renewcommand{\arraystretch}{1.15}
\resizebox{\columnwidth}{!}{%
\begin{tabular}{lccc}
\toprule
Prosodic Feature F0 & MUStARD Cohen $d$ & CMMA Cohen $d$ & Cross-cultural Shift \\
\midrule
f0\_mean & -0.180 & -0.030 & 6$\times$ Attenuation \\
f0\_var & +0.130 & +0.040 & 3$\times$ Attenuation \\
loudness & +0.340 & -0.090 & Direction Inversion \\
spectralFlux & +0.260 & -0.100 & Direction Inversion \\
tempo & -0.150 & +0.020 & Vector Decay \\
HNR & 0.000 & +0.040 & Statistically Flat \\
\midrule
Average absolute $|d|$ & 0.177 & 0.053 & 3.3$\times$ Magnitude Drop \\
\bottomrule
\end{tabular}%
}
\end{table}

\begin{table}[h]
\centering
\caption{Cross-cultural variation of high-level Omni perception probe outputs.}
\label{tab:cross_cultural_high_level}
\large
\setlength{\tabcolsep}{3pt}
\renewcommand{\arraystretch}{1.15}
\resizebox{\columnwidth}{!}{%
\begin{tabular}{lccc}
\toprule
Omni Perception Probe Head & MUStARD $\Delta$pp & CMMA $\Delta$pp & Cross-cultural Fate \\
\midrule
P3 Tone-text Mismatch & +8.100 & +22.300 & 2.75$\times$ Amplification \\
P2 Hostile Style Probe & +6.200 & +18.300 & 2.95$\times$ Amplification \\
P2 Sincere Style Probe & -12.200 & -12.300 & Cross-lingual Parity \\
P1 Intensity Strong Probe & -3.900 & +3.800 & Inverted Informative \\
\bottomrule
\end{tabular}%
}
\end{table}


The data in Tables~\ref{tab:cross_cultural_low_level} and~\ref{tab:cross_cultural_high_level} reveal a clear architectural decoupling. Low-level physical prosody collapses sharply in the transition from English to Chinese speech. The average absolute effect size drops by $3.3\times$, and loudness and spectral flux even invert direction. High-level cognitive features captured by the auditory probes, by contrast, remain robust and undergo substantial cross-cultural amplification. In particular, the tone-text mismatch probe $P_3$ shifts from 8.1 to 22.3 percentage points.

These patterns converge with established findings in psycholinguistics. Cheang and Pell~\cite{cheang2009acoustic} demonstrate that English sarcasm utilises a lowered mean $F_0$ with restricted variability and elevated spectral noise. Conversely, Cantonese sarcasm displays an elevated mean $F_0$ within a narrow range. They further report that reliable cross-language sarcasm identification depends on native experience. Li et al.~\cite{li2020role} characterise Mandarin sarcasm as voice-quality-driven and marked primarily by creakier phonation, a profile that contrasts with the spectrally noisy English case. Prosodic statistics fitted on English speech therefore cannot transfer cleanly to Mandarin corpora. 

The cross-cultural amplification of probe $P_3$ aligns with Echoic Mention Theory~\cite{wilson2006pragmatics}, under which irony marks a cognitive contrast between an echoed expectation and the actual utterance. In tonal languages, lexical tones occupy the $F_0$ channel for semantic identity, which restricts the bandwidth available for affective prosodic exaggeration. We hypothesise that the pragmatic contrast then projects preferentially onto the lexical-prosodic interface, rather than relying on pitch contour alone. This specific interface matches the precise signal targeted by the high-level probe $P_3$. The dual-stream integration of ALFR successfully captures this language-invariant cognitive anchor.

\subsubsection{Prompt-Language Transferability}
\label{sec:prompt_language_transferability}
We evaluate the cross-lingual robustness of our zero-shot prompt design by testing English calibration templates verbatim on the Chinese CMMA corpus against localised translations.

\begin{table}[h]
\centering
\caption{Prompt-language transferability evaluations on the CMMA corpus reporting Macro-F1. Best performance within each backbone family across all prompt settings is highlighted in bold.}
\label{tab:cross_cultural_prompts}
\large
\setlength{\tabcolsep}{5pt}
\renewcommand{\arraystretch}{1.15}
\resizebox{\columnwidth}{!}{%
\begin{tabular}{lcccccc}
\toprule
& \multicolumn{3}{c}{English Prompt Templates} & \multicolumn{3}{c}{Chinese Prompt Templates} \\ 
\cmidrule(lr){2-4}\cmidrule(lr){5-7}
LLM Backbone Family & DZS & +BiCAL & +ALFR & DZS & +BiCAL & +ALFR \\
\midrule
Mistral-Small-Instruct-2409   & 0.527 & 0.572   & 0.591  & 0.468 & 0.508   & \textbf{0.606} \\
Qwen2.5-32B-Instruct         & 0.577 & 0.599   & 0.594  & 0.497 & 0.598   & \textbf{0.622} \\
Gemma-2-27B-it          & 0.273 & 0.286   & \textbf{0.607}  & 0.140 & 0.187   & 0.601 \\
Qwen2.5-7B          & 0.269 & 0.279   & \textbf{0.586}  & 0.280 & 0.298   & 0.560 \\
Llama-3.1-8B-Instruct        & 0.144 & 0.193   & \textbf{0.575}  & 0.115 & 0.138   & 0.541 \\
DeepSeek-V4-Pro     & 0.542 & 0.592   & \textbf{0.619}  & 0.477 & 0.551   & 0.603 \\
GPT-5.4             & 0.545 & 0.604   & \textbf{0.627}  & 0.362 & 0.504   & 0.621 \\
\bottomrule
\end{tabular}%
}
\end{table}

We examine whether the BiCAL templates require explicit translation on the Chinese CMMA corpus. Holding the execution methodology fixed, replacing the five English templates with Chinese counterparts produces near-identical Stage 1 results on some backbones. For instance, Qwen2.5-32B-Instruct shifts by a marginal 0.001 in Macro-F1 score. Conversely, this modification triggers a sharp collapse on other architectures, where the performance drop tracks the Chinese-language alignment of the model rather than its raw parameter capacity. Under Chinese prompting, the Macro-F1 score of Gemma-2-27B-it declines by an absolute margin of 0.099 ($0.286 \to 0.187$). The same degradation appears even at the frontier. The Macro-F1 score of GPT-5.4 falls by 0.100 ($0.604 \to 0.504$), whereas the Chinese-developed DeepSeek-V4-Pro is markedly more robust, experiencing an absolute decrease of only 0.041 ($0.592 \to 0.551$). Native Chinese sarcasm queries thus trigger a more severe positivity bias on backbones weakly aligned to Chinese, irrespective of scale.

The Stage 2 ALFR rescue narrows this prompt-language gap substantially across all seven backbones. The Macro-F1 score of Gemma-2-27B-it converges to within 0.006 between the two prompt conditions, while Llama-3.1-8B-Instruct converges to within 0.034 in Macro-F1 score. The two API backbones exhibit a similar convergence pattern. The initial 0.100 collapse of GPT-5.4 shrinks to a residual gap of 0.006 in Macro-F1 score ($0.627 \to 0.621$), while the cross-prompt gap for DeepSeek-V4-Pro is restricted to 0.016 ($0.619 \to 0.603$). ALFR consistently reduces prompt-language sensitivity but does not fully eliminate it, leaving a small residual gap of roughly 0.006 to 0.020 in Macro-F1 score. We adopt English templates uniformly in our main evaluations to maintain cross-cultural transferability. The Chinese prompt template for CMMA is available in Appendix~\ref{app:zh_prompts}.

\subsection{Explainability Analysis}
\label{sec:interpretability_analysis}
\begin{figure}[t]
\centering
\includegraphics[width=\linewidth]{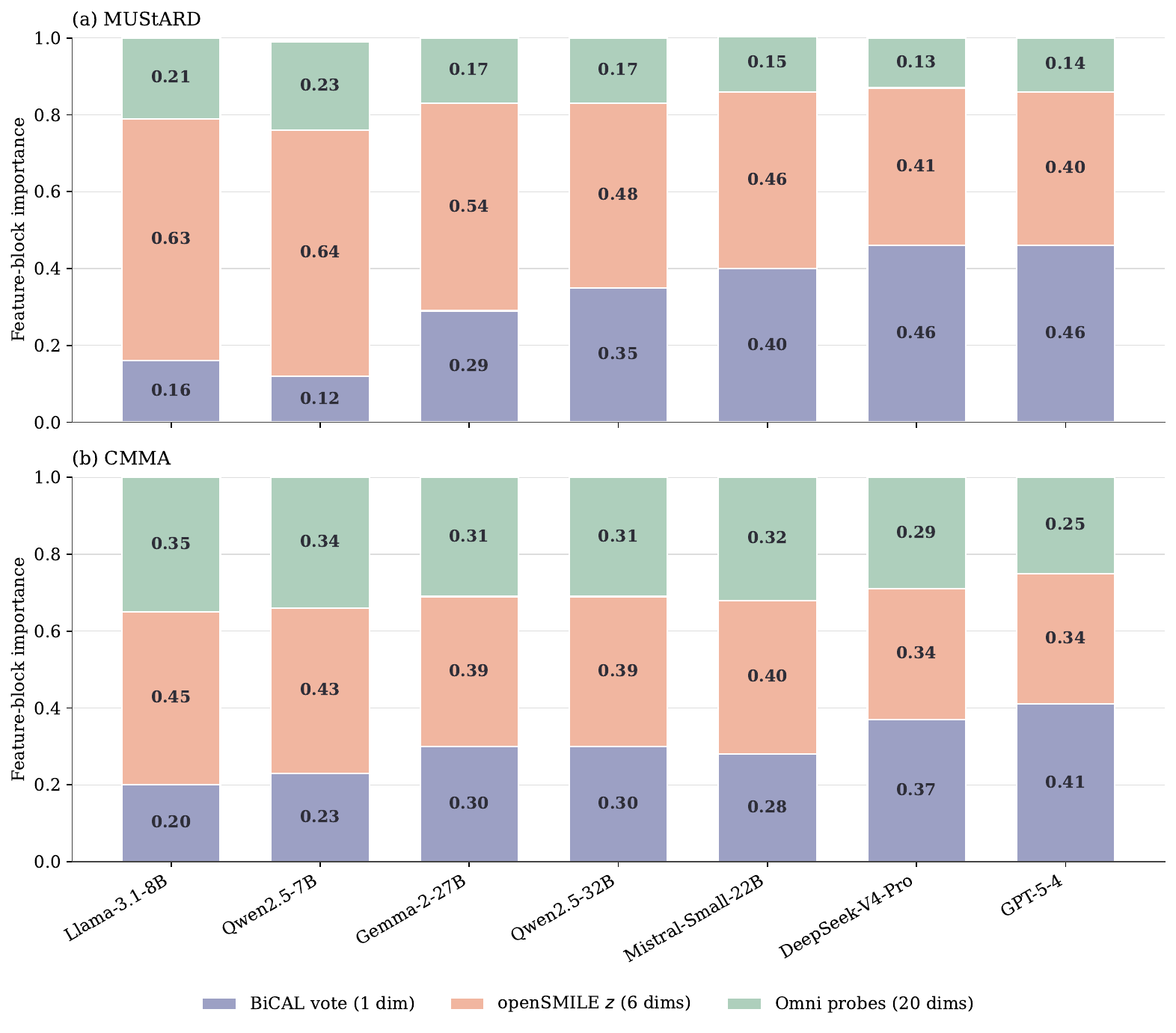}
\caption{Random Forest impurity-based feature importance~\cite{breiman2001random} allocated to the three ALFR feature blocks (BiCAL vote, openSMILE prosodic $z$, Omni perception probes) per LLM backbone, on (a)~MUStARD and (b)~CMMA. Importances are summed within each block and normalised to one per LLM. Random Forest is used for cross-backbone consistency only.}
\label{fig:feature_importance}
\end{figure}

We conduct a post-hoc explainability analysis of the internal routing dynamics of ALFR. Figure~\ref{fig:feature_importance} presents impurity-based feature importance allocations across LLM backbones, computed with Random Forest~\cite{breiman2001random} for cross-backbone consistency. 

On the English MUStARD dataset, where the horizontal axis orders backbones by baseline textual strength, the BiCAL vote importance rises steadily as the text backbone strengthens, from 0.160 on Llama-3.1-8B-Instruct to 0.400 on Mistral-Small-Instruct-2409, while the low-level openSMILE $z$ importance contracts from 0.630 to 0.460. As stronger language models produce more reliable calibration signals, the late-fusion classifier shifts weight onto this calibrated text channel and away from raw prosody functionals.

A different allocation profile emerges on the Chinese CMMA benchmark, where the vote importance stays between 0.200 and 0.300 across all architectures while the high-level Omni perception probe block rises sharply, reaching a mean importance of 0.326 on CMMA. This level scales significantly above the 0.186 mean observed on MUStARD. This routing transition represents a 1.75$\times$ amplification. Section~\ref{sec:acoustic_feature_decoupling} quantifies this cross-cultural audio signal asymmetry at the feature level, and the present analysis tracks the same phenomenon at the routing level. Both converge on one mechanism: low-level physical prosody fails to transfer cross-culturally. The late-fusion head compensates by re-weighting toward high-level perceptual probes. This routing behaviour provides the language-invariant cognitive anchor that sustains cross-lingual classification stability.

\section{Conclusion}
\label{sec:conclusion}
This paper presented CHARM, a two-stage framework for multimodal sarcasm detection that decouples pragmatic processing into a textual bidirectional calibration layer and an acoustically grounded late-fusion rescue module. Across seven LLM backbones, the calibration layer consistently mitigated systemic positive-class over-prediction, achieving highly significant Stouffer meta-analysis scores of $Z = 5.85$ on MUStARD and $Z = 19.29$ on CMMA. The acoustic late-fusion module then delivered substantial gains for pragmatically compromised small networks, compressing the inter-model performance gap by $2.5\times$ on MUStARD and $7.9\times$ on CMMA across all seven backbones, with cumulative significance of $Z = 13.89$ and $Z = 34.64$. Finally, feature-block decompositions exposed a cross-cultural prosodic decoupling phenomenon. Low-level physical prosody collapsed across language environments, whereas high-level audio-language perception probes amplified on Chinese data. This robust perceptual channel provided the language-invariant cognitive anchor needed to sustain stable cross-lingual classification.

Several limitations bound the scope of this study. The framework relies on shallow classifier routing heads optimised on validation splits rather than a fully zero-shot pipeline, and its high-level features depend on a single audio foundation model without testing alternative architecture families. Additionally, our late-fusion layer relies on feature concatenation, which may fail to capture complex non-linear interactions across different sarcasm subtypes, such as semantic-prosodic incongruity or purely acoustic stylistic modulations. The cross-cultural validation covers only one English and one Chinese benchmark, omitting other tonal-language families, and the diagnostics operate purely at the behavioural output level without white-box probing of internal hidden states. Finally, although we contextualise CHARM against supervised multimodal systems, these comparisons remain indicative rather than strictly head-to-head, since the methods differ in training regime, modality, and evaluation protocol. Future work will integrate the late-fusion features into a unified auxiliary supervised fine-tuning loop, explore advanced cross-modal attention or tensor fusion mechanisms to dynamically model distinct sarcasm sub-phenomena, and transfer the calibration mechanism to adjacent pragmatic tasks. Furthermore, we intend to align these machine-derived attributions with native cross-cultural human annotations to build more granular explainability frameworks.

\appendices
\setcounter{section}{0}
\counterwithin{table}{section}
\counterwithin{figure}{section}

\section{Late-Fusion Classifier Hyperparameters}
\label{sec:classifier_hyperparameters}

Table~\ref{tab:classifier_hyperparams} lists the configuration of the ten candidates evaluated by the late-fusion sweep. On the imbalanced CMMA benchmark, class re-weighting is applied as documented in the rightmost column, whereas on MUStARD that column remains inactive.

\begin{table}[h]
\centering
\caption{Late-fusion classifier hyperparameters. ``Scaled'' indicates the estimator is wrapped in a Pipeline with a preceding StandardScaler. $\eta$ denotes learning rate, and $n_{\text{est}}$ denotes the number of base estimators. Unspecified parameters retain library defaults under scikit-learn 1.6.1, xgboost 3.2.0, and lightgbm 4.6.0.}
\label{tab:classifier_hyperparams}
\large
\setlength{\tabcolsep}{3pt}
\renewcommand{\arraystretch}{1.18}
\resizebox{\columnwidth}{!}{%
\begin{tabular}{l l l}
\toprule
Classifier   & Hyperparameters & CMMA Re-weighting Pool \\
\midrule
LR           & $C{=}1.0$, \texttt{max\_iter}${=}2000$, scaled              & \texttt{class\_weight=`balanced'} \\
RF           & $n_{\text{est}}{=}200$, \texttt{max\_depth}${=}6$            & \texttt{class\_weight=`balanced'} \\
GB           & $n_{\text{est}}{=}200$, \texttt{max\_depth}${=}3$, $\eta{=}0.05$ & --- \\
AdaBoost     & $n_{\text{est}}{=}200$, $\eta{=}0.5$                         & --- \\
SVM (RBF)    & $C{=}1.0$, $\gamma{=}$\texttt{scale}, scaled                 & \texttt{class\_weight=`balanced'} \\
MLP          & layers${=}(64, 32)$, \texttt{max\_iter}${=}500$, early stopping, scaled & --- \\
XGBoost      & $n_{\text{est}}{=}200$, \texttt{max\_depth}${=}3$, $\eta{=}0.05$ & \texttt{scale\_pos\_weight}${=}7.8$ \\
LightGBM     & $n_{\text{est}}{=}200$, \texttt{max\_depth}${=}3$, $\eta{=}0.05$ & \texttt{class\_weight=`balanced'} \\
Stacking     & base${=}\{$LR, RF, GB$\}$, meta${=}$LR                       & Inherited \\
Soft Voting  & estimators${=}\{$LR, RF, GB$\}$, \texttt{voting=`soft'}      & Inherited \\
\bottomrule
\end{tabular}%
}
\end{table}

\section{Full 7-LLM $\times$ 10-Classifier Sweep Matrix}
\label{app:classifier_sweep}

Tables~\ref{tab:sweep_mustard} and~\ref{tab:sweep_cmma} report the full late-fusion sweep underlying the Best CLF column of Table~\ref{tab:main_results}. Each cell reports Macro-F1 and Accuracy averaged over 5-fold, 5-seed cross-validation; the optimal classifier per language backbone family is highlighted in boldface.
\begin{table*}[t]
\centering
\caption{Full 7-LLM $\times$ 10-classifier sweep on the MUStARD benchmark with sample count equal to 690. Each cell reports the Macro-F1 and Accuracy values. The optimal classifier per architecture is highlighted in boldface.}
\label{tab:sweep_mustard}
\small
\setlength{\tabcolsep}{6pt}
\renewcommand{\arraystretch}{1.15}
\resizebox{\linewidth}{!}{%
\begin{tabular}{l ccccccc}
\toprule
Classifier & Mistral-Small-Instruct-2409 & Qwen2.5-32B-Instruct & Gemma-2-27B-it  & Qwen2.5-7B & Llama-3.1-8B-Instruct & DeepSeek-V4-Pro & GPT-5.4 \\
\midrule
LR          & \textbf{0.729/0.730} & \textbf{0.729/0.730} & 0.705/0.705 & 0.619/0.620 & 0.627/0.629 & 0.780/0.780 & 0.782/0.782 \\
RF          & 0.727/0.728          & 0.718/0.719          & 0.718/0.718 & \textbf{0.635/0.636} & 0.622/0.624 & 0.783/0.783 & 0.783/0.783 \\
GB          & 0.719/0.720          & 0.698/0.699          & 0.707/0.708 & 0.616/0.617          & 0.637/0.638 & 0.767/0.768 & 0.761/0.761 \\
AdaBoost    & 0.723/0.724          & 0.728/0.729          & \textbf{0.721/0.725} & 0.610/0.612          & 0.620/0.622 & 0.784/0.785 & 0.772/0.772 \\
SVM (RBF)   & 0.714/0.715          & 0.709/0.709          & 0.696/0.696 & 0.616/0.618          & 0.622/0.624 & 0.769/0.769 & 0.754/0.754 \\
MLP         & 0.704/0.705          & 0.701/0.701          & 0.665/0.666 & 0.598/0.600          & 0.612/0.614 & 0.748/0.749 & 0.747/0.747 \\
XGBoost     & 0.715/0.715          & 0.702/0.702          & 0.707/0.708 & 0.609/0.610          & 0.632/0.633 & 0.766/0.767 & 0.755/0.756 \\
LightGBM    & 0.715/0.715          & 0.710/0.711          & 0.701/0.702 & 0.620/0.622          & \textbf{0.635/0.641} & 0.758/0.759 & 0.752/0.752 \\
Stacking    & 0.726/0.726          & 0.728/0.728          & 0.721/0.722 & 0.628/0.629          & 0.635/0.635 & 0.783/0.783 & \textbf{0.787/0.787} \\
Soft Voting & 0.725/0.726          & 0.723/0.723          & 0.722/0.723 & 0.632/0.634          & 0.630/0.632 & \textbf{0.784/0.785} & 0.784/0.784 \\
\bottomrule
\end{tabular}%
}
\end{table*}

\begin{table*}[t]
\centering
\caption{Full 7-LLM $\times$ 10-classifier sweep on the CMMA benchmark under English prompt conditions. Each cell reports the Macro-F1 and Accuracy values. The optimal classifier per architecture is highlighted in boldface with balanced class re-weighting constraints applied.}
\label{tab:sweep_cmma}
\small
\setlength{\tabcolsep}{6pt}
\renewcommand{\arraystretch}{1.15}
\resizebox{\linewidth}{!}{%
\begin{tabular}{l ccccccc}
\toprule
Classifier & Mistral-Small-Instruct-2409 & Qwen2.5-32B-Instruct & Gemma-2-27B-it  & Qwen2.5-7B & Llama-3.1-8B-Instruct & DeepSeek-V4-Pro & GPT-5.4 \\
\midrule
LR & 0.535/0.652 & 0.569/0.703 & 0.540/0.660 & 0.524/0.637 & 0.515/0.620 & 0.585/0.714 & 0.594/0.723 \\
RF & 0.557/0.714 & 0.577/0.741 & 0.562/0.725 & 0.559/0.720 & 0.544/0.710 & 0.586/0.741 & 0.600/0.750 \\
GB & 0.494/0.885 & 0.493/0.883 & 0.486/0.882 & 0.485/0.884 & 0.475/0.884 & 0.504/0.882 & 0.505/0.882 \\
AdaBoost & 0.470/0.886 & 0.470/0.886 & 0.470/0.886 & 0.470/0.886 & 0.470/0.886 & 0.472/0.887 & 0.471/0.887 \\
SVM (RBF) & 0.525/0.642 & 0.557/0.687 & 0.533/0.657 & 0.525/0.646 & 0.519/0.639 & 0.566/0.690 & 0.591/0.727 \\
MLP & 0.471/0.886 & 0.472/0.886 & 0.470/0.886 & 0.471/0.886 & 0.468/0.886 & 0.471/0.886 & 0.480/0.886 \\
XGBoost & 0.551/0.690 & 0.563/0.713 & 0.554/0.698 & 0.550/0.693 & 0.538/0.678 & 0.577/0.718 & 0.590/0.727 \\
LightGBM & 0.552/0.693 & 0.566/0.717 & 0.555/0.700 & 0.550/0.695 & 0.539/0.690 & 0.579/0.721 & 0.588/0.728 \\
Stacking & 0.532/0.644 & 0.559/0.687 & 0.535/0.650 & 0.518/0.625 & 0.509/0.610 & 0.578/0.701 & 0.588/0.713 \\
Soft Voting & \textbf{0.591/0.823} & \textbf{0.594/0.824} & \textbf{0.607/0.840} & \textbf{0.586/0.836} & \textbf{0.575/0.831} & \textbf{0.619/0.829} & \textbf{0.627/0.819} \\
\bottomrule
\end{tabular}%
}
\end{table*}

\section{Random Paraphrase Control Templates}
\label{app:randpara}

\begin{figure}[h]
\centering

\definecolor{paraMain}{RGB}{90,110,120}   
\definecolor{paraSys}{RGB}{120,120,120}   

\begin{tcolorbox}[enhanced, width=\linewidth,
    colback=white, colframe=black!70, boxrule=0.8pt, arc=3pt,
    left=8pt, right=8pt, top=7pt, bottom=8pt,
    title={\centering\bfseries Random Paraphrase Control\\[1pt]
           \normalfont\small\itshape 5 neutral variants, all at zero charge},
    coltitle=white, fonttitle=\bfseries, colbacktitle=black!70,
    halign title=center]

\begin{tcolorbox}[enhanced, colback=paraSys!8, colframe=paraSys,
    boxrule=0.5pt, arc=2pt, left=7pt, right=7pt, top=4pt, bottom=4pt,
    title style={fill=paraSys}, fonttitle=\bfseries, coltitle=white,
    title={Shared System Prompt \hfill \normalfont\footnotesize\itshape $=$ BiCAL zero-charge}]
  \footnotesize
  ``You answer questions about short dialogue snippets. Reply with a single word.''
\end{tcolorbox}

\vspace{5pt}

\begin{tcolorbox}[enhanced, colback=paraMain, colframe=paraMain,
    boxrule=0pt, arc=2pt, left=7pt, right=7pt, top=2.5pt, bottom=2.5pt,
    coltext=white, before skip=2pt, after skip=5pt]
  \footnotesize\bfseries Variant-Specific User Tails
  \hfill\normalfont\itshape differing wording in \textbf{bold}
\end{tcolorbox}

\newcommand{\pararow}[2]{
  \begin{tcolorbox}[enhanced, colback=paraMain!5, colframe=paraMain,
      boxrule=0.5pt, arc=2pt, left=7pt, right=7pt, top=3pt, bottom=4pt,
      borderline west={3pt}{0pt}{paraMain},
      before skip=4pt, after skip=4pt]
    \footnotesize
    {\ttfamily\bfseries\color{paraMain!50!black} #1}\quad #2
  \end{tcolorbox}}

\pararow{[paraphrase\_1]}{Is \texttt{\{speaker\}} \textbf{being sarcastic}? Answer Yes or No.}
\pararow{[paraphrase\_2]}{Does \texttt{\{speaker\}} \textbf{use sarcasm}? Answer Yes or No.}
\pararow{[paraphrase\_3]}{Is \texttt{\{speaker\}}'s \textbf{remark sarcastic}? Answer Yes or No.}
\pararow{[paraphrase\_4]}{Is \texttt{\{speaker\}} \textbf{expressing sarcasm}? Answer Yes or No.}
\pararow{[paraphrase\_5]}{Would you \textbf{classify} \texttt{\{speaker\}}'s \textbf{statement as sarcastic}? Answer Yes or No.}

\end{tcolorbox}
\caption{The five neutral-paraphrase templates used by the Random Paraphrase Control of
Sec.~\ref{sec:random_paraphrase_control}. All variants share the exact system prompt of BiCAL's zero-charge variant
(shown once at the top); only the user-tail phrasing varies, with the differing wording
highlighted in \textbf{bold}. Dynamic input slots are set in a typewriter typeface.}
\label{fig:randpara_templates}
\end{figure}
Figure~\ref{fig:randpara_templates} reports the five neutral paraphrase templates used by the Random Paraphrase Control framework. All five share an identical system prompt matching the neutral configuration of BiCAL and differ only in the surface phrasing of the user query. None contains stance-forcing language or explicit expert or skeptic role specifications; consequently, each configuration occupies a stable position near the zero-charge centre of the bias axis.

The sampling profile under which these control templates are executed is identical to the main BiCAL module. The generation temperature is locked at 0.5 and the top-$p$ threshold at 0.95, across three random seeds (42, 43, and 44). We collect three samples per seed-template combination, accumulating fifteen evaluation votes per instance, and the final consensus prediction follows the same majority-rule equation $\hat{y} = \mathbb{1}[v > 7]$ deployed in the standard BiCAL calibration runs.

\section{Chinese Prompt Templates on CMMA}
\label{app:zh_prompts}

\begin{figure}[h]
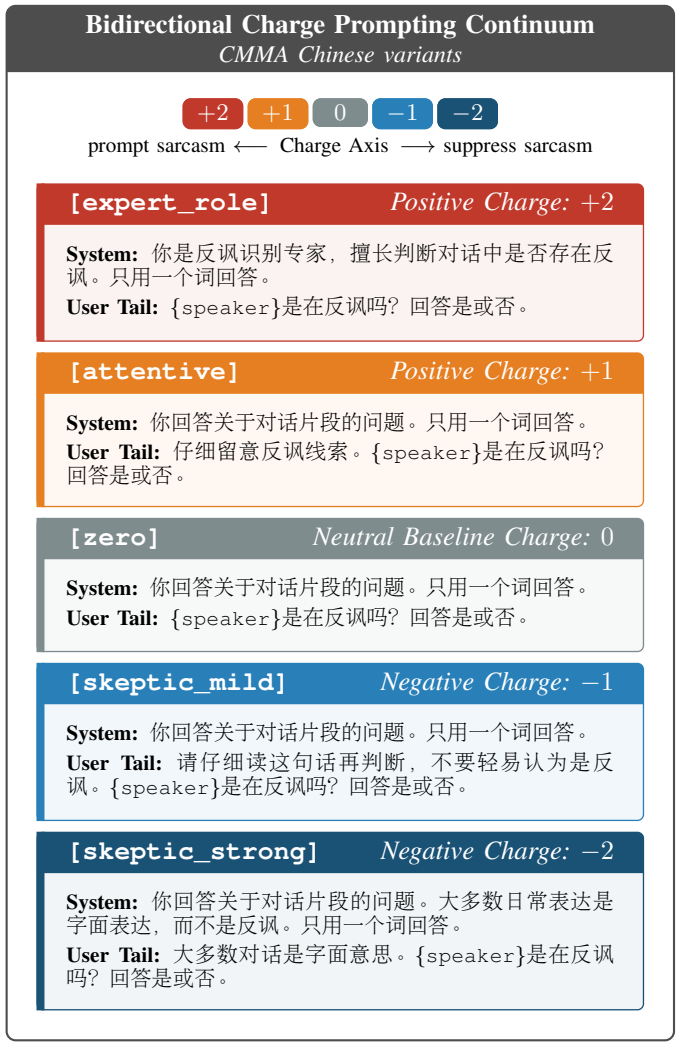

\centering

\newcommand{\zh}[1]{\begin{CJK*}{UTF8}{gbsn}#1\end{CJK*}}

\newcommand{\chgbadge}[2]{%
  \tcbox[on line, colback=#1, colframe=#1, coltext=white,
         boxsep=1.5pt, left=4pt, right=4pt, top=1pt, bottom=1pt,
         arc=3pt, boxrule=0pt, nobeforeafter]{\bfseries\small #2}}

\newcommand{\zhvariant}[5]{%
  \begin{tcolorbox}[enhanced, colback=#1!6, colframe=#1,
      boxrule=0.5pt, arc=2pt, left=8pt, right=8pt, top=4pt, bottom=5pt,
      borderline west={3pt}{0pt}{#1},
      title style={fill=#1}, fonttitle=\bfseries, coltitle=white,
      before skip=4pt, after skip=4pt,
      title={\ttfamily #2 \hfill \normalfont\itshape #3}]
    \footnotesize
    \textbf{System:}\, #4\\[2pt]
    \textbf{User Tail:}\, #5
  \end{tcolorbox}}

\begin{tcolorbox}[enhanced, width=\linewidth,
    colback=white, colframe=black!70, boxrule=0.8pt, arc=3pt,
    left=8pt, right=8pt, top=7pt, bottom=8pt,
    title={\centering\bfseries Bidirectional Charge Prompting Continuum\\[1pt]
           \normalfont\small\itshape CMMA Chinese variants},
    coltitle=white, fonttitle=\bfseries, colbacktitle=black!70,
    halign title=center]

{\centering
\chgbadge{posStrong}{$+2$}\,\chgbadge{posMild}{$+1$}\,%
\chgbadge{neutral}{$\;0\;$}\,\chgbadge{negMild}{$-1$}\,%
\chgbadge{negStrong}{$-2$}\\[3pt]
\footnotesize prompt sarcasm $\longleftarrow$\, Charge Axis \,$\longrightarrow$ suppress sarcasm\par}
\vspace{6pt}

\zhvariant{posStrong}{[expert\_role]}{Positive Charge: $+2$}
  {\zh{你是\textbf{反讽识别专家}，擅长判断对话中是否存在反讽。只用一个词回答。}}
  {\zh{\texttt{\{speaker\}}是在反讽吗？回答 是 或 否。}}

\zhvariant{posMild}{[attentive]}{Positive Charge: $+1$}
  {\zh{你回答关于对话片段的问题。只用一个词回答。}}
  {\zh{\textbf{仔细留意反讽线索。}\texttt{\{speaker\}}是在反讽吗？回答 是 或 否。}}

\zhvariant{neutral}{[zero]}{Neutral Baseline Charge: $0$}
  {\zh{你回答关于对话片段的问题。只用一个词回答。}}
  {\zh{\texttt{\{speaker\}}是在反讽吗？回答 是 或 否。}}

\zhvariant{negMild}{[skeptic\_mild]}{Negative Charge: $-1$}
  {\zh{你回答关于对话片段的问题。只用一个词回答。}}
  {\zh{\textbf{请仔细读这句话再判断，不要轻易认为是反讽。}\texttt{\{speaker\}}是在反讽吗？回答 是 或 否。}}

\zhvariant{negStrong}{[skeptic\_strong]}{Negative Charge: $-2$}
  {\zh{你回答关于对话片段的问题。\textbf{大多数日常表达是字面表达，而不是反讽。}只用一个词回答。}}
  {\zh{\textbf{大多数对话是字面意思。}\texttt{\{speaker\}}是在反讽吗？回答 是 或 否。}}

\end{tcolorbox}
\caption{The Chinese symmetrical bidirectional charge prompting continuum within the
BiCAL module on CMMA. Variants are colour-coded along the symmetric charge axis, from
strongly prompting sarcasm ($+2$) to strongly suppressing it ($-2$); colour encodes
charge polarity (warm $=$ positive, cool $=$ negative), and the charge-bearing wording
is highlighted in \textbf{bold}. Dynamic slots are set in a typewriter typeface.}
\label{fig:cmma_zh_templates}
\end{figure}

Figure~\ref{fig:cmma_zh_templates} presents the native Chinese prompt templates evaluated on the CMMA corpus. 
The five prompt configurations mirror the symmetric structure of the English BiCAL templates, operating along the same $[-2, +2]$ charge axis. All sampling parameters are identical to the core evaluation profile: a temperature of 0.5 and a top-$p$ threshold of 0.95 across three random seeds.

\section{Model Card}
\label{app:model_cards}

Table~\ref{tab:model_card} lists the foundation models invoked by the CHARM framework, documenting their HuggingFace identifiers, parameter scales, licences, and operational roles. All models are evaluated as-is, without parameter updates. No fine-tuning or adapter adjustments are applied to any underlying backbone.

\begin{table*}[t]
\centering
\caption{Model card for all foundation models invoked by CHARM. Open-weight parameter counts follow the totals reported in official release notes and are loaded in native \texttt{bfloat16}; commercial or frontier APIs list parameter counts based on official technical briefs where available. The audio language model executes via its native multimodal interface following the official technical report recipes.}
\label{tab:model_card}
\small
\setlength{\tabcolsep}{5pt}
\renewcommand{\arraystretch}{1.18}
\resizebox{\linewidth}{!}{%
\begin{tabular}{lllll}
\toprule
Model Name & HuggingFace Identifier / API & Param. & License & Role in CHARM \\

\midrule
Qwen2.5-32B-Instruct  & \texttt{Qwen/Qwen2.5-32B-Instruct}               & 32.5B & Apache 2.0 & Text backbone (BiCAL) \\
Mistral-Small-Instruct-2409     & \texttt{mistralai/Mistral-Small-Instruct-2409}   & 22.0B & Mistral Research License & Text backbone (BiCAL) \\
Gemma-2-27B-it            & \texttt{google/Gemma-2-27B-it}                   & 27.2B & Gemma Terms of Use & Text backbone (BiCAL) \\
Qwen2.5-7B-Instruct   & \texttt{Qwen/Qwen2.5-7B-Instruct}                 & 7.6B  & Apache 2.0 & Text backbone (BiCAL) \\
Llama-3.1-8B-Instruct & \texttt{meta-llama/Llama-3.1-8B-Instruct}         & 8.0B  & Llama-3 Community License & Text backbone (BiCAL) \\
Qwen2.5-Omni-7B       & \texttt{Qwen/Qwen2.5-Omni-7B}                     & 7.0B  & Apache 2.0 & Audio probe extractor (ALFR $\Psi$) \\

DeepSeek-V4-Pro       & \texttt{deepseek-v4-pro} (API)                   & 1.6T & MIT & Text backbone (BiCAL) \\
GPT-5.4               & \texttt{gpt-5.4} (API)                           & Undisclosed & Proprietary API & Text backbone (BiCAL) \\
\bottomrule
\end{tabular}%
}
\end{table*}

\paragraph*{Hardware and inference configuration.}
We load open-source backbones in their native \texttt{bfloat16} precision on a single NVIDIA A100 80GB GPU. DeepSeek-V4-Pro and GPT-5.4 are instead queried through their respective inference APIs and incur no local GPU cost. Inference follows the decoding specifications documented in our experimental setup. The direct zero-shot baseline uses greedy decoding with $\mathtt{do\_sample}$ set to false, whereas the BiCAL and random-paraphrase variants deploy stochastic sampling profiles instead. This configuration sets the temperature $T$ to 0.5, the $\text{top-}p$ threshold to 0.95, and the sample allocation $N$ to 3 per template-seed combination. The audio probing pipeline maintains greedy decoding on Qwen2.5-Omni-7B. No model parameter undergoes fine-tuning at any framework stage.

\clearpage

\bibliographystyle{IEEEtran}
\bibliography{sample-base}



\end{document}